\documentclass[11pt, a4paper]{article}

\usepackage[force]{feynmp-auto}
\usepackage{jheppub,multirow,relsize,slashed,textpos}
\usepackage{wasysym}
\usepackage{marvosym}
\usepackage{tikz}
\usepackage{braket} 
\usepackage{physics} 
\usepackage{placeins}
\usepackage{booktabs} 
\usepackage{accents} 
\usepackage{multirow} 
\usepackage{float}

\newcommand{\refeq}[1]{Eq.~(\ref{#1})}

\newcommand{\reffig}[1]{Fig.~\ref{#1}}

\newcommand{\refsec}[1]{Section~\ref{#1}}
\newcommand{\refapp}[1]{Appendix~\ref{#1}}
\newcommand{\reftab}[1]{Table~\ref{#1}}

\newcommand*{\pbar}[1]{\accentset{(-)}{#1}}

\definecolor{light-gray}{gray}{0.65}

\newcounter{CommentCount}
\graphicspath{{figs/}}

\title{Neutrino Trident Scattering at Near Detectors}

\author[a]{Peter Ballett}
\author[a]{\!\!, Matheus Hostert}
\author[a]{\!\!, Silvia Pascoli}
\author[b,c]{\!\!, Yuber F. Perez-Gonzalez}
\author[b]{\!\!, Zahra Tabrizi}
\author[b]{and Renata Zukanovich Funchal}

\affiliation[a]{Institute for Particle Physics Phenomenology, Department of
Physics, Durham University,\\ 
	South Road, Durham DH1 3LE, United Kingdom}

\affiliation[b]{Departamento de F\'isica Matem\'atica, Instituto de
  F\'isica, Universidade de S\~ao Paulo,\\
  R.\ do Mat\~ao 1371, CEP.\ 05508-090, S\~ao Paulo, Brazil}

\affiliation[c]{ICTP South American Institute for Fundamental Research
  \& Instituto de F\'isica Te\'orica,\\
  Universidade Estadual Paulista, Rua Dr.\ Bento T.\ Ferraz 271,
  CEP.\ 01140-070, S\~ao Paulo, Brazil}

\emailAdd{peter.ballett@durham.ac.uk}
\emailAdd{matheus.hostert@durham.ac.uk}
\emailAdd{silvia.pascoli@durham.ac.uk}
\emailAdd{yfperezg@if.usp.br}
\emailAdd{ztabrizi@if.usp.br}
\emailAdd{zukanov@if.usp.br}

\preprint{IPPP/18/64}

\abstract{Neutrino trident scattering is a rare Standard Model process where a charged-lepton pair is produced in neutrino-nucleus scattering. To date, only the dimuon final-state has been observed, with around 100 total events, while the other channels are as yet unexplored.
In this work, we compute the trident production cross section by performing a complete four-body phase space calculation for different hadronic targets. This provides a correct estimate both of the coherent and the diffractive contributions to these cross sections, but also allows us to address certain inconsistencies in the literature related to the use of the Equivalent Photon Approximation in this context. We show that this approximation can give a reasonable estimate only for the production of dimuon final-states in coherent scattering, being inadmissible for all other cases considered.
We provide estimates of the number and distribution of trident events at several current and future near detector facilities subjected to intense neutrino beams from accelerators: five liquid-argon detectors (SBND, $\mu$BooNE, ICARUS, DUNE and $\nu$STORM), the iron detector of T2K (INGRID) and three detectors made of composite material (MINOS, NO$\nu$A and MINER$\nu$A). 
We find that for many experiments, trident measurements are an attainable goal and a valuable addition to their near detector physics programme.
}

\begin{document} 

\maketitle


\section{Introduction}
\label{sec:intro}

The Standard Model (SM) has been confronted with a variety of experimental data and has so far emerged as an impressive phenomenological description of nature, except in the neutrino sector. The observation of neutrino flavour oscillations by solar, atmospheric, reactor and accelerator neutrino experiments over the last 50 years has revealed the existence of neutrino mass and flavour mixing, making necessary the first significant extension of the SM.

The precise determination of the neutrino mixing parameters as well as the search for the neutrino mass ordering and leptonic CP violation drive both present and future accelerator neutrino experiments. To accomplish these tasks, these experiments rely on state-of-the-art near detectors, made of heavy materials, located a few hundred meters downstream of the neutrino source and subjected to a high intensity beam. Their main purpose is to ensure high precision measurements at a far detector by reducing the systematic uncertainties related to neutrino fluxes, charged-current (CC) and neutral-current (NC) cross sections and backgrounds.  
The high beam luminosity they are subjected to (about $10^{21}$ protons on target) and their relatively large fiducial mass of high-$Z$ materials (typically 100~ton) make these detectors ideal places to investigate rare neutrino-nucleus interactions ($\sigma\lesssim 10^{-44}$~cm${}^2$), such as neutrino trident scattering. 

Trident events are processes predicted by the SM as the result of (anti)neutrino-nucleus scattering with the production of a charged lepton pair \cite{Czyz:1964zz,Lovseth:1971vv,Fujikawa:1971nx,Brown:1971qr,Koike:1971tu}, $\pbar{\nu}_{\alpha}+{\cal{H}} \to \pbar{\nu}_{\alpha \,{\rm{or}} \, \kappa(\beta)} + \ell_{\beta}^- + \ell_\kappa^+ +{\cal{H}}$, $\{\alpha,\beta,\kappa\}\in \{e,\mu,\tau\}$\footnote{Throughout the manuscript we will consider ${\alpha,\beta, \kappa}$ as flavour indexes.} where $\cal{H}$ denotes a hadronic target. Depending on the (anti)neutrino and charged lepton flavours in the final-state, the process will be mediated by the $Z^0$ boson, $W$ boson or both. Coherent interactions between (anti)neutrinos and the atomic nuclei are expected to dominate these processes as long as the momentum transferred $Q$ is significantly smaller than the inverse of the nuclear size \cite{Czyz:1964zz}. For larger momentum transfers diffractive and deep-inelastic scattering become increasingly relevant \cite{Magill:2016hgc}.
Although this process exists for all combinations of same-flavour or mixed flavour charged-lepton final-states, to this day only the $\nu_\mu$-induced dimuon mode, $\pbar{\nu}_\mu + {\cal{H}} \to \pbar{\nu}_\mu  + \mu^+ + \mu^- + {\cal{H}}$, has been observed. The first measurement of this trident signal performed by CHARM II~\cite{Geiregat:1990gz} is also the one with the largest statistics: 55 signal events in a beam of neutrinos and antineutrinos with $\langle E_\nu \rangle \approx 20$ GeV. Other measurements by CCFR~\cite{Mishra:1991bv} and NuTeV~\cite{Adams:1998yf} at larger energies soon followed.

As the measurement of trident events may provide a sensitive test of the weak sector~\cite{Brown:1973ih} as well as placing constraints on physics beyond the SM~\cite{Mishra:1991bv,Gaidaenko:2000hg,Altmannshofer:2014pba,Kaneta:2016uyt,Ge2017,Magill:2017mps,Falkowski:2018dmy} it is relevant to investigate how to probe the other modes. This was re\-cog\-ni\-zed by the authors of Ref.~\cite{Magill:2016hgc} who calculated the cross sections for trident production in all possible flavour combinations and estimated the number of events expected for the DUNE and SHiP experiments. They used the Equivalent Photon Approximation (EPA)~\cite{Belusevic:1987cw} to compute the cross section in the coherent and diffractive regimes of the scattering. The EPA, however, is known to breakdown for final state electrons~\cite{Kozhushner:1962aa, Shabalin:1963aa, Czyz:1964zz} leading, as we will demonstrate here, to an overestimation of the cross section that in some cases is by more than 200\%. 

In this work, we present a unified treatment of the coherent and diffractive trident calculation beyond the EPA for all modes. We then compute the number and distribution of events expected in each mode at various near detectors, devoting particular attention to the case of liquid argon (LAr) detectors, as they are expected to lead the field of precision neutrino scattering measurements over the next few decades thanks to their excellent tracking and calorimetry capabilities. Finally, we address the likely backgrounds that may hinder these experimental searches --- a question that we believe to be of utmost importance given the rarity of the process, and one that has been omitted in earlier sensitivity studies \cite{Magill:2016hgc,Altmannshofer:2014pba}.

This paper is organized as follows. In Sec.~\ref{sec:xsec}, we explain how to correctly calculate the trident SM cross sections, comparing our results to the EPA and explicitly showing the breakdown of this approximation. In Sec.~\ref{sec:LAr}, we discuss the trident event rates and kinematic distributions at the near detectors of several present and future neutrino oscillation experiments based on LAr technology: the three detectors of the Short-Baseline Neutrino (SBN) Program at Fermilab~\cite{SBNproposal} and the near detector for the long-baseline Deep Underground Neutrino Experiment (DUNE)~\cite{Acciarri:2016ooe,DUNECDRvolII}, also located at Fermilab. We also consider the potential gains from an optimistic future facility: a 100 t LAr detector subject to the novel low-systematics neutrino beam of the Neutrinos from STORed Muons ($\nu$STORM) project~\cite{Soler:2015ada,nuSTORM2017}. 
We discuss the sources of background events at these facilities, providing a GENIE-level analysis \cite{Andreopoulos2009} of how to reduce these backgrounds and assessing the impact they are expected to have on the trident measurement. 
In Sec.~\ref{sec:others}, we discuss other near detectors that use more conventional technologies: the Interactive Neutrino GRID (INGRID)~\cite{Abe:2011xv,Abe:2015biq,Abe:2016fic,Abe:2016tez}, the on-axis iron near detector for T2K at J-PARC, as well as three detectors at the Neutrino at the Main Injector (NuMI) beamline at Fermilab, the one for the Main INjector ExpeRiment $\nu$-A (MINER$\nu$A)~\cite{Altinok:2017xua,MINERvA:2017} and the near detectors for the Main Injector Oscillation Search (MINOS)~\cite{Adamson:2014pgc,AlpernBoehm} and the Numi Off-axis $\nu_e$ Appearance (NO$\nu$A) experiment~\cite{Wang:Biao,sanchez_mayly_2018_1286758}. 
Finally, in Sec.~\ref{sec:conc} we  present our last remarks and conclusions.

\section{Trident Production Cross Section}\label{sec:xsec}

In this section we consider neutrino trident production in the SM, defined as the process where a (anti)neutrino scattering off a hadronic system ${\cal H}$ produces a pair of same-flavour or mixed flavour charged leptons 
\begin{equation}
\pbar{\nu}_{\alpha}(p_1) \,+\, {\cal H}(P) \,\to\, \pbar{\nu}_{\alpha\, {\rm or}\, \kappa(\beta)} (p_2) \,+\, \ell_\beta^- (p_4)  \,+\, \ell_\kappa^+ (p_3) \,+\, {\cal H}(P^\prime),
\label{eq:indices}
\end{equation}
where $\beta (\kappa)$ corresponds to the flavour index of the negative (positive) charged lepton in both neutrino and antineutrino cases. 
Neutrino trident scattering can be divided into three regimes depending on the nature of the hadronic target: coherent, diffractive and deep inelastic, when the neutrino scatters off the nuclei, nucleons and quarks, respectively. At the energies relevant for neutrino oscillation experiments, the deep inelastic scattering contribution amounts at most to 1\% of the total trident production cross section \cite{Magill:2016hgc} and we will not consider it further.

The cross section for trident production has been calculated before in the literature, both in the context of the $V-A$ theory~\cite{Czyz:1964zz,Lovseth:1971vv,Fujikawa:1971nx} and in the SM~\cite{Brown:1973ih}, while the EPA treatment was developed in Refs.~\cite{Kozhushner:1962aa,Shabalin:1963aa,Belusevic:1987cw}. Most calculations have focused on the coherent channels \cite{Czyz:1964zz,Lovseth:1971vv,Fujikawa:1971nx,Brown:1973ih,Belusevic:1987cw} but the diffractive process has been considered in \cite{Czyz:1964zz,Lovseth:1971vv}. More recently, calculations using the EPA have been performed for coherent scattering with a dimuon final-state \cite{Altmannshofer:2014pba}, and for all combinations of hadronic targets and flavours of final-states in \cite{Magill:2016hgc}. While the EPA is expected to agree reasonably well with the full calculation for coherent channels with dimuon final-states, the assumptions of this approximation are invalid for the coherent process with electrons in the final-state \cite{Kozhushner:1962aa,Shabalin:1963aa,Czyz:1964zz}. 
For this reason, we perform the full $2\to 4$ calculation without the EPA in a manner applicable to any hadronic target, following a similar approach to Refs.~\cite{Czyz:1964zz,Lovseth:1971vv}. Our treatment of the cross section allows us to quantitatively assess the breakdown of the EPA in both coherent and diffractive channels for all final-state flavours, an issue we come back to in Sec.~\ref{sec:EPAbreakdown}.

We write the total cross section for neutrino trident production off a nucleus ${\cal N}$ with $Z$ protons and $(A-Z)$ neutrons as the sum
\begin{equation}
\sigma_\mathrm{\nu {\cal N}} = \sigma_\mathrm{\nu c} +\sigma_\mathrm{\nu d}\, ,
\end{equation}
where $\sigma_\mathrm{\nu c}$ ($\sigma_\mathrm{\nu d}$) is the coherent (diffractive) part of the cross section. 
%
\unitlength = 0.9mm
\begin{figure}[t]
\centering\includegraphics[width=\textwidth]{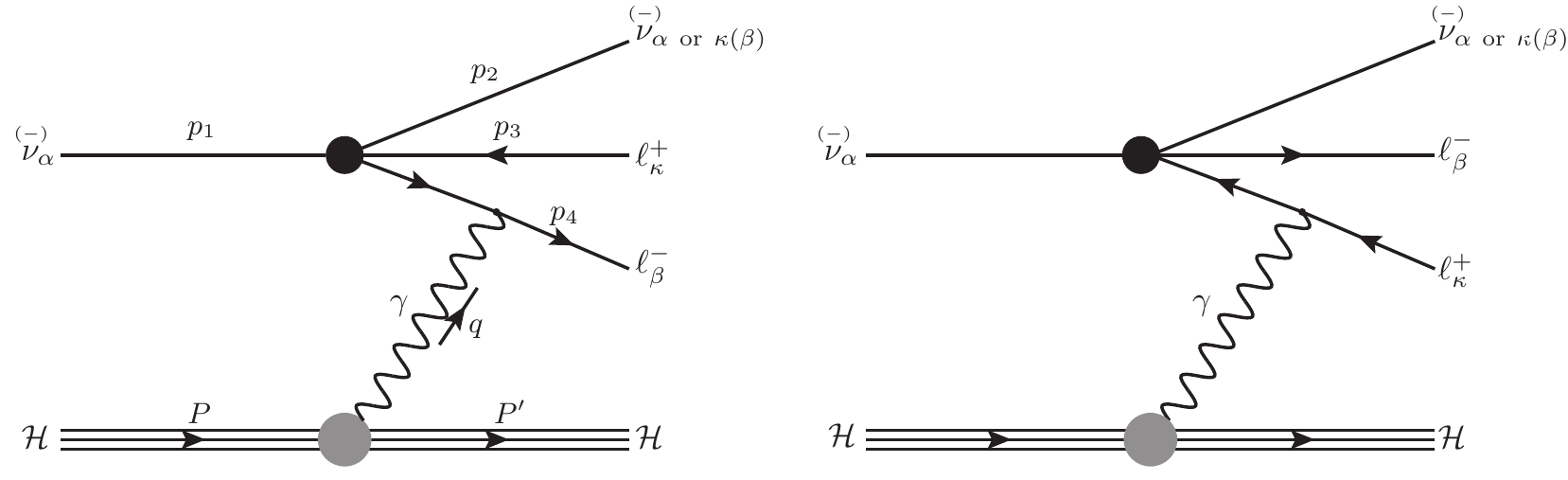}
\caption{Diagrams contributing to the neutrino trident process in the four-point interaction limit of the Standard Model.  
\label{fig:Tdiagrams}}
\end{figure}

%
The relevant diagrams for these processes in the coherent or diffractive 
regimes involve the boson $Z^0$, $W$ or both mediators, depending on the particular mode. In the four-point interaction limit, depicted in \reffig{fig:Tdiagrams}, these reduce to only two contributions, one where the photon couples to the negatively and one to the positively charged lepton. 
In Table \ref{tab:tridentmodes} we present the processes that will be considered in this work as well as the SM contributions present in each. Although our formalism applies also to processes with final-state $\tau$ leptons, the increased threshold makes them irrelevant for the experiments of interest in this study and we do not consider them further.
The trident amplitude for a coherent (${\rm X=c}$) or diffractive (${\rm X=d}$) scattering regime can be written as
\begin{equation}
i \mathcal{M} = \mathrm{L}^\mu (\{p_i\},q) \, \frac{-ig_{\mu \nu}}{q^2} \, \mathrm{H}_{\rm X}^{\nu}(P,P^{\prime})\, ,
\end{equation}
where $\{p_i\}=\{p_2,p_3,p_4\}$ is the set of outgoing leptonic momenta. $ \mathrm{L}^\mu (\{p_i\},q)$ is the total leptonic amplitude 
\begin{align}
\mathrm{L}^\mu & \equiv - \frac{ie G_F}{\sqrt{2}}[\bar{u}(p_2)\gamma^\tau(1-\gamma_5)u(p_1)] \times 
\bar{u}(p_4)\left[\gamma_\tau(V_{\alpha\beta\kappa}-A_{\alpha\beta\kappa}\gamma_5)\frac{1}{(\slashed{q}-\slashed{p}_3-m_3)}\gamma^\mu \right . \nonumber \\ 
& \left . + \gamma^\mu \frac{1}{(\slashed{p}_4-\slashed{q}-m_4)} \gamma_\tau (V_{\alpha\beta\kappa}-A_{\alpha\beta\kappa}\gamma_5) \right] v(p_3)\, ,
\label{eq:Lmu}
\end{align}
and $\mathrm{H}_{\rm X}^{\nu}(P,P^{\prime})$ is the total hadronic amplitude
\begin{align}
H_{\rm X}^\nu  &\equiv \langle {\cal H}(P) \vert J_\mathrm{E.M.}^\nu (q^2)\vert {\cal H}(P^\prime)\rangle\, ,
\label{eq:Hmu}
\end{align}
with  $q \equiv P - P^\prime$ denoting the transferred momentum, $m_3$ ($m_4$) the positively (negatively) charged lepton mass, $V_{\alpha\beta\kappa}(A_{\alpha\beta\kappa})\equiv g_{V}^{\beta}(g_A^{\beta})\delta_{\beta\kappa}+\delta_{\alpha\beta} \,(\beta=\alpha \, \mathrm{or} \; \kappa)$ the vector (axial) couplings, depending on the channel and have labels in accordance to Eq. (\ref{eq:indices}), and ${J}^\nu_{\rm{E.M.}} (q^2)$ the electromagnetic current for the hadronic system ${\cal H}$ (a nucleus or a nucleon).
\begin{table}[t]
\begin{center}
\begin{tabular}{|cc|}
\toprule\toprule
\bf (Anti)Neutrino &  \bf SM Contributions \\
\midrule\midrule
$\pbar{\nu}_\mu\, {\cal H} \to \pbar{\nu}_\mu\, \mu^- \mu^+\,  {\cal H}$  & CC + NC \\
$\pbar{\nu}_\mu\, {\cal H} \to \pbar{\nu}_e\,  e^\pm \mu^\mp\, {\cal H}$  & CC\\
$\pbar{\nu}_\mu\, {\cal H} \to \pbar{\nu}_\mu\,  e^- e^+\, {\cal H}$ & NC\\
$\pbar{\nu}_e\, {\cal H} \to \pbar{\nu}_e\,  e^- e^+\, {\cal H}$ &  CC + NC \\
$\pbar{\nu}_e\, {\cal H} \to \pbar{\nu}_\mu\,  \mu^\pm e^\mp\, {\cal H}$ & CC \\
$\pbar{\nu}_e\, {\cal H} \to \pbar{\nu}_e\,  \mu^- \mu^+\, {\cal H}$ & NC \\
\bottomrule
\bottomrule
\end{tabular}
\end{center}
\caption{\label{tab:tridentmodes} (Anti)Neutrino trident processes considered in this paper.}
\end{table}

We can write the differential cross section as
\begin{align}
\frac{\dd^2 \sigma_{\nu  {\rm X}}}{\dd Q^2 \dd \hat{s}}= \frac{1}{32  \pi^2(s-M_{\cal H}^2)^2}\frac{\mathrm{H}_{\rm X}^{\mu\nu}\mathrm{L}_{\mu\nu}}{Q^4}\, ,
\end{align}
where $s = (p_1 + P)^2$, $\hat{s} \equiv 2\,(p_1 \vdot q)$, $Q^2 = -q^2$ and $M_{\cal H}$ is the mass of the hadronic target. We have also introduced the hadronic tensor $\mathrm{H}_{\rm X}^{\mu \nu}$
\begin{align}
\mathrm{H}_{\rm X}^{\mu\nu} &\equiv \overline{\sum_{\rm{spins}}}  \left(\mathrm{H}_{\rm X}^\mu\right)^* \mathrm{H}_{\rm X}^\nu.
\end{align}
The two scattering regimes in which the hadronic tensor is computed will be discussed in more detail in Sec.~\ref{sec:had}. The leptonic tensor, $\mathrm{L}^{\mu \nu}$, integrated over the phase space of the three final-state leptons, $\dd^{3} \Pi \left(p_1 + q; \{p_i\}\right)$, and merely summed over final and initial spins is given by
\begin{equation}
\mathrm{L}^{\mu \nu} (p_1, q) \equiv  \int \dd ^{3} \Pi \left(p_1 + q; \{p_i\}\right) \left( \sum_{\rm{spins}} \left(  \mathrm{L}^\mu \right)^*  \mathrm{L}^\nu  \right)\, .
\label{eq:Lmunu}
\end{equation}
We can use $\mathrm{L}^{\mu \nu}$ to define two scalar functions, one related to the longitudinal ($\mathrm{L}_{\mathrm{L}}$) and the other to the transverse ($\mathrm{L}_{\mathrm{T}}$) polarization of the exchanged photon
\begin{equation}
\mathrm{L}_{\mathrm{T}} = -\frac{1}{2}\left( g^{\mu \nu} - \frac{4Q^2}{\hat{s}^2} p_1^\mu p_1^\nu \right) \mathrm{L}_{\mu \nu}, \quad \mathrm{and} \quad \mathrm{L_{L}} =  \frac{4Q^2}{\hat{s}^2} p_1^\mu p_1^\nu \mathrm{L}_{\mu \nu}\, .\label{eq:LT_LL}
\end{equation}
This allows us to write the differential cross section as a sum of a longitudinal and a transverse contribution \cite{Hand:1963bb} as follows
\begin{align}
\frac{\dd^2 \sigma_{\nu  {\rm X}}}{\dd Q^2 \dd \hat{s}} &= \frac{1}{32 \pi^2} \frac{1}{\hat{s}\,Q^2} \left [ h_{\rm X}^\mathrm{T}(Q^2, \hat{s}) \, \sigma^\mathrm{T}_{\nu \gamma} (Q^2, \hat{s}) + h_{\rm X}^\mathrm{L}(Q^2, \hat{s}) \, \sigma^\mathrm{L}_{\nu \gamma} (Q^2, \hat{s}) \right] \, ,\label{eq:full_diff_xsec}
\end{align}
where we have defined two functions for the flux of longitudinal and transverse virtual photons 
\begin{subequations}
\label{eq:splitting_function}
\begin{align}
h_{\rm X}^\mathrm{T}(Q^2, \hat{s})  &\equiv \frac{2}{(E_\nu M_{\cal H})^2} \left[ p_{1 \mu} p_{1\nu} -\frac{\hat{s}^2}{4 Q^2} \, g_{\mu \nu} \right]\mathrm{H}_{\rm X}^{\mu \nu} , \quad \text{and} \\ \quad
h_{\rm X}^\mathrm{L}(Q^2, \hat{s})  &\equiv \frac{1}{(E_\nu M_{\cal H})^2} \, p_{1\mu} p_{1\nu}\, \mathrm{H}_{\rm X}^{\mu \nu}\, ,
\end{align}
\end{subequations}
and two leptonic neutrino-photon cross sections associated with them\footnote{Note that we include a factor of $1/2$ in $\sigma^\mathrm{T}_{\nu \gamma}$ to match the polarization averaging of the on-shell cross section: $\sigma_{\nu \gamma}^{\rm on-shell} = \frac{1}{2 \hat{s}} \left( \overline{\sum}_r (\epsilon_r^\mu)^* \epsilon^\nu_r \, {\rm L}_{\mu\nu} \right) \big\vert_{Q^2=0} = \frac{1}{4 \hat{s}} \left( - g^{\mu\nu} L_{\mu\nu}\right) \big\vert_{Q^2=0} = \frac{{\rm L_T}}{2 \hat{s}}\big\vert_{Q^2=0} = \sigma_{\nu\gamma}^\text{T}(0,\hat{s})$.}
\begin{equation}
\sigma^\mathrm{T}_{\nu \gamma} (Q^2, \hat{s})  = \frac{\mathrm{L_T}}{2 \hat{s}}\, , \quad \mathrm{and} \quad
\sigma^\mathrm{L}_{\nu \gamma} (Q^2, \hat{s})  = \frac{\mathrm{L_L}}{\hat{s}}\, .
\end{equation}
The kinematically allowed region in the $(Q^2,\hat{s})$ plane can be obtained by considering the full four-body phase space, as in~\cite{Czyz:1964zz,Lovseth:1971vv,Fujikawa:1971nx}. The limits for such physical region are given by
\begin{subequations}\label{eq:qslimts}
	\begin{align}
		Q_{\rm min}^2&=\frac{M_{\cal H} \hat{s}^2}{2E_\nu(2E_\nu M_{\cal H}-\hat{s})},&\  
        Q_{\rm max}^2&=\hat{s}-m_L^2,\label{eq:qlimts}\\
        \hat{s}_{\rm min}&=\frac{E_\nu}{2E_\nu + M_{\cal H}}\left[m_L^2+2E_\nu M_{\cal H} -\Delta\right]&\  
        \hat{s}_{\rm max}&=\frac{E_\nu}{2E_\nu + M_{\cal H}}\left[m_L^2+2E_\nu M_{\cal H} +\Delta\right],\label{eq:shatlimts}
	\end{align}
\end{subequations}
with $m_L\equiv m_3+m_4$, and
\begin{align*}
	\Delta \equiv \sqrt{(2E_\nu M_{\cal H}-m_L^2)^2-4M_{\cal H}^2 m_L^2}\,.
\end{align*}
Let us emphasize that \refeq{eq:full_diff_xsec} is an exact decomposition, and does not rely on any approximation of the process. In the following section, we will show how to calculate the flux functions $h_{\rm X}^\mathrm{T}$ and $h_{\rm X}^\mathrm{L}$ from Eq.~\ref{eq:splitting_function} in different scattering regimes. The total cross section for the process can then be computed by finding $\sigma^\mathrm{L}_{\nu \gamma}$ and $\sigma^\mathrm{T}_{\nu \gamma}$ from Eqs.~(\ref{eq:Lmu}), (\ref{eq:Lmunu}) and (\ref{eq:LT_LL}) and integrating over all allowed values of $Q^2$ and $\hat{s}$. Note that $\sigma^\mathrm{L}_{\nu \gamma}$ and $\sigma^\mathrm{T}_{\nu \gamma}$ are universal functions for a given leptonic process and need only to be computed once.

\subsection{Hadronic Scattering Regimes}\label{sec:had}

Depending on the magnitude of the virtuality of the photon, $Q = \sqrt{-q^2}$, the hadronic current can contribute in different ways to the trident process. Thus, given the decomposition in \refeq{eq:full_diff_xsec}, the change in the hadronic treatment translates to computing the flux factors $h_{\rm X}^\mathrm{T}$ and $h_{\rm X}^\mathrm{L}$ for each scattering regime.  From those flux factors, $\sigma_{\nu\mathrm{c}}$ and $\sigma_{\nu\mathrm{d}}$ can be calculated.

\subsubsection{Coherent Regime (${\rm H}^{\mu \nu}_{\rm c}$)}

In the coherent scattering regime the incoming neutrino interacts with the whole nucleus without resolving its substructure. For this to occur frequently, we need small values of $Q$. Despite the relatively large neutrino energies in contemporary neutrino beams, this is still allowed for trident.

In this regime, the hadronic tensor $\mathrm{H}^{\mu\nu}_\mathrm{c}$ for a ground state spin-zero nucleus of charge $Z e$ can be written in terms 
of the nuclear electromagnetic form factor $F(Q^2)$, discussed in more detail in Appendix~\ref{app:formfactors}, as
\begin{equation}
\mathrm{H}^{\mu \nu}_\mathrm{c} =  4Z^2 e^2 \left| F (Q^2)\right|^2 \left(P^\mu - \frac{q^\mu}{2}\right) \left(P^\nu - \frac{q^\nu}{2}\right).
\end{equation}
From Eq.~\ref{eq:splitting_function}, we find that the transverse and longitudinal flux functions for the coherent regime are
\begin{subequations}\label{eq:hcoh}
\begin{align}
h^\mathrm{T}_\mathrm{c}(Q^2, \hat{s})  &=  8 Z^2 e^2   \left(1 - \frac{\hat{s}}{2E_\nu M} - \frac{\hat{s}^2}{4 E_\nu^2 Q^2}\,\right) |F (Q^2)|^2\, , \\
h^\mathrm{L}_\mathrm{c}(Q^2, \hat{s})  &=  4 Z^2 e^2  \left(1 - \frac{\hat{s}}{4E_\nu M}\right)^2 |F (Q^2)|^2\, ,
\end{align}
\end{subequations}
where $E_\nu$ is the energy of the incoming neutrino and $M$ is the nuclear mass. For a fixed value of $\hat{s}$ in the physical region, the $h^{\rm T}_{\rm c}$ flux function becomes zero at $Q_{\rm min}$ while the longitudinal component does not. This different behaviour can be seen explicitly in their definitions, Eqs. \eqref{eq:hcoh}, as the terms in the parenthesis in $h^{\rm T}_{\rm c}$ cancel each other at $Q_{\rm min}$. This does not occur for $h^{\rm L}_{\rm c}$ since the physical values of $\hat{s}$ are always smaller than $E_\nu M$ in this hadronic regime. 
Due to this fact, $Q_{\rm min}$, which according to  Eq.\ \eqref{eq:qlimts} depends on  both the  neutrino energy and target material, can be approximated to
\begin{align*}
	Q_{\rm min} \approx \frac{\hat{s}}{2E_{\nu}},
\end{align*}
which only depends on the incoming neutrino energy. On the other hand, as $Q$ becomes large, the flux functions $h^{T,L}$ become quite similar, $h^{\rm T}_{\rm c}\approx 2 h^{\rm L}_{\rm c}$, and favour small values of $\hat{s}$. After some critical value of the virtuality $Q$, $h^{\rm T, L}_{\rm c}$ become negligible due to the nuclear form factor. The $Q$ value at which this happens depends on the target material, but not on the incoming neutrino energy. For instance, in the case of an Ar target
the flux functions basically vanish for $Q\gtrsim 250$ MeV.

\begin{figure}[t]
\centering
\includegraphics[width=\textwidth]{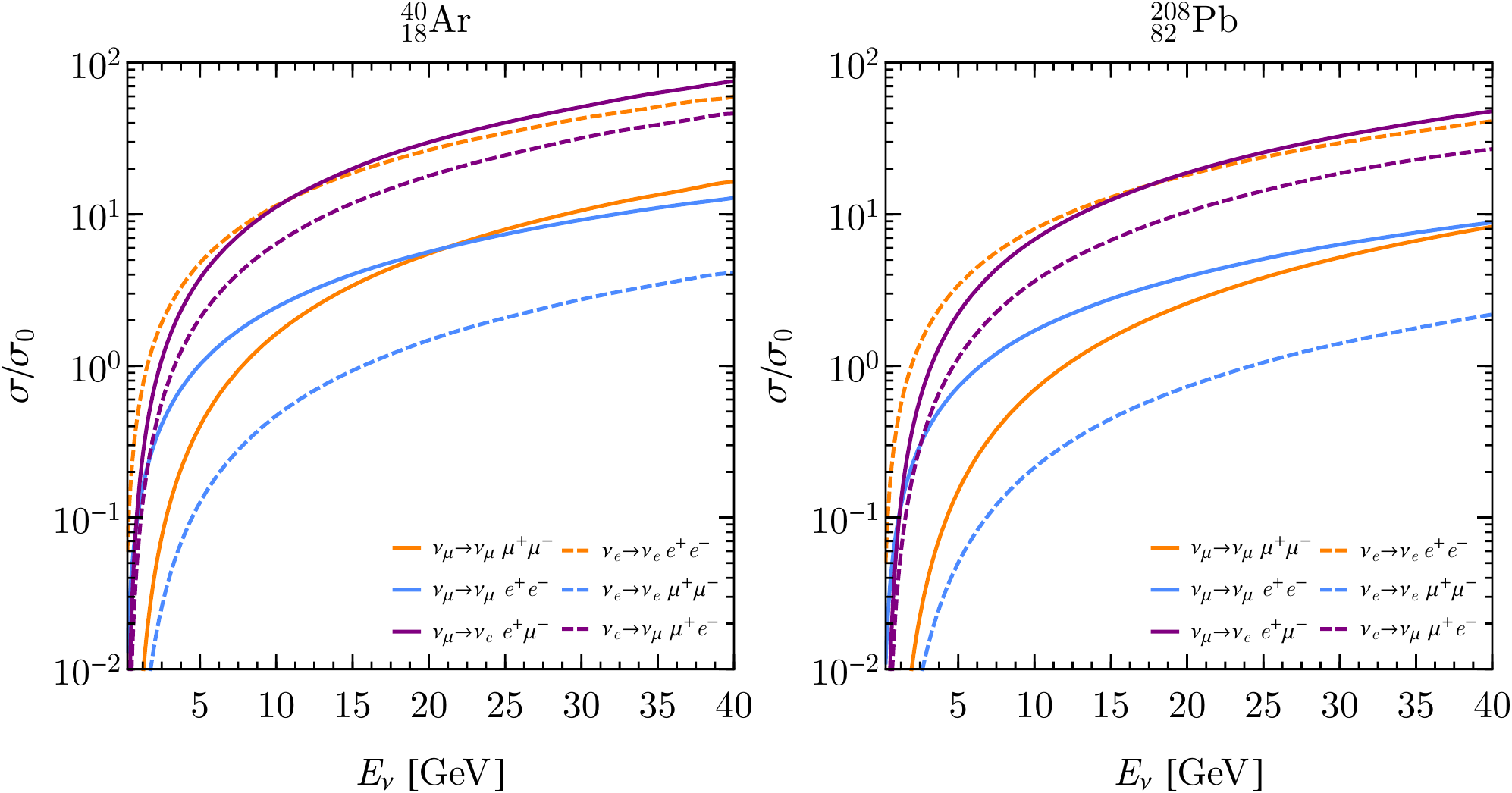}
\caption{Cross sections for coherent neutrino trident production on $^{40}$Ar (left) and  $^{208}$Pb (right) normalized to $\sigma_0 =  Z^2\, 10^{-44}$ cm$^2$. The full (dashed) lines correspond to the scattering of an incoming $\nu_\mu$ ($\nu_e$) produced by the NC (light-blue), CC (purple), and CC+NC (orange) SM interactions. \label{fig:coh_xsec}}
\end{figure}

The final cross sections for coherent neutrino trident production on Argon can be seen in \reffig{fig:coh_xsec}. Despite thresholds being important for the behaviour of these cross sections for GeV neutrino energies, we can see that mixed channels quickly become the most important due to their CC nature. At large energies one can then rank the cross sections from largest to smallest as CC, CC+NC, and NC only channels. Nevertheless, one must be aware of the fact that the cross sections are dominated by low $Q^2$ even at large energies, leading to large effects due to the final-state lepton masses as discussed in \cite{Magill:2016hgc}.

\subsubsection{Diffractive Regime ($\mathrm{H}^{\mu\nu}_\mathrm{d}$)}

At larger $Q^2$, the neutrino interacts with the individual nucleons of the nucleus. In this diffractive regime $\mathrm{H}^{\mu\nu}_\mathrm{d}$ is given by the sum of the contributions of the two types of nucleons: protons ($\mathrm{N=p}$) and neutrons ($\mathrm{N=n}$), so
\begin{equation}
\mathrm{H}^{\mu \nu}_\mathrm{d} (P, P^\prime) = Z\, \mathrm{H}^{\mu \nu}_\mathrm{p} (P, P^\prime)+
(A-Z)\,\mathrm{H}^{\mu \nu}_\mathrm{n}(P, P^\prime)\, ,
\end{equation}
where each $\mathrm{H}^{\mu \nu}_\mathrm{N}$ is the square of the matrix element of the nucleon electromagnetic current summed over final and averaged over initial spins. Neglecting second class currents, the matrix elements take the form
\begin{equation}
\bra{\mathrm{N}(P^\prime)} {J}^\mu_{\rm{E.M.}} (Q^2) \ket{\mathrm{N} (P) } = e \, \overline{u}_\mathrm{N} (P^\prime) \left[ \gamma^\mu F^\mathrm{N}_1(Q^2) - i \frac{\sigma^{\mu \nu} q_{\nu}}{2 M_{\rm N}} F^\mathrm{N}_2(Q^2) \right] u_\mathrm{N}(P)\, ,
\end{equation}
with $F^\mathrm{N}_{1,2}(Q^2)$ the Dirac and Pauli form factors, respectively. The hadronic tensors are then given by \cite{Kniehl:1990iv}
\begin{equation}
\mathrm{H}^{\mu \nu}_\mathrm{N} = e^2 \left[ 4 \, H_1^\mathrm{N}(Q^2) \left(P^\mu - \frac{q^\mu}{2}\right)\left(P^\nu - \frac{q^\nu}{2}\right) - H_2^\mathrm{N}(Q^2) \left( Q^2 g^{\mu \nu} + q^\mu q^\nu \right) \right]\, ,
\end{equation}
where the  $H_1^\mathrm{N}(Q^2)$ and $H_2^\mathrm{N}(Q^2)$ form factors, functions of $F^\mathrm{N}_{1,2}(Q^2)$, are given in Appendix~\ref{app:formfactors}. The flux functions in the diffractive regime can then be calculated as
\begin{subequations}\label{eq:dcoh}
\begin{align}
h^\mathrm{T}_\mathrm{N}(Q^2, \hat{s})  &=  8 \, e^2 \left[ \left(1 - \frac{\hat{s}}{2E_\nu M_{\rm N}} - \frac{\hat{s}^2}{4 E_\nu^2 Q^2 }\,\right) H_1^\mathrm{N}(Q^2) + \frac{\hat{s}^2}{8E_\nu^2 M_{\rm N}^2}  H_2^\mathrm{N}(Q^2)\right ]\, ,\label{eq:hTdiff}\\
h^\mathrm{L}_\mathrm{N}(Q^2, \hat{s})  &=  4 e^2 \, \left[ \left(1-\frac{\hat{s}}{4 E_\nu M_{\rm N}} \right)^2 H_1^\mathrm{N}(Q^2)  - \frac{\hat{s}^2}{16 E_\nu^2 M_{\rm N}^2} H_2^\mathrm{N}(Q^2)\right]\, .\label{eq:hLdiff}
\end{align}
\end{subequations}
In the case of the proton, the flux functions $h^{\rm T, L}_{\rm p}$ have some unique features given the presence of both electric and magnetic contributions. Specifically, the transverse function is non-zero at $Q=Q_{\rm min}$ for a fixed $\hat{s}$, due to the additional term proportional to $H_2^{\rm p}$. Indeed, for large values of $\hat{s}$, the $H_2^{\rm p}$ term dominates the transverse function. An opposite behaviour occurs for the longitudinal component. There, the $H_1^{\rm p}$ term dominates over the second term for all physical values of $\hat{s}$, $Q$, and for any incoming neutrino energy. On the other hand, the flux functions of the neutron, which have only 
the magnetic moment contribution, have somewhat different characteristics. While 
$h^{\rm T}_{\rm n}$ behaves similarly to $h^{\rm T}_{\rm p}$, that is, it is dominated by the second term for large values of $\hat{s}$, $h^{\rm L}_{\rm n}$ is zero at $Q_{\rm min}$ due to the exact cancellation between the $H_{1,2}^{\rm n}$ terms. This cancellation is not evident from Eq.\ ~\eqref{eq:hLdiff}; however, simplifying the longitudinal component for the neutron case, one finds
\begin{align*}
	h^\mathrm{L}_\mathrm{n}(Q^2, \hat{s})  &=4 e^2 \left(1+\frac{Q^2}{4M_{\rm n}^2}\right)\frac{Q^2}{4 M_{\rm N}^2}\left( 1 - \frac{\hat{s}}{2 E_\nu M_{\rm N}} - \frac{\hat{s}^2}{4 E_\nu^2 Q^2} \right) \left| F^\mathrm{n}_2(Q^2) \right|^2,
\end{align*}
which is zero for $Q=Q_{\rm min}$. Also, this shows why $h^{\rm L}_{\rm p}$ does not 
vanish at $Q_{\rm min}$ since there we have the additional contribution of the electric component. 

When the neutrino interacts with an individual nucleon inside the nucleus, one must be aware of the nuclear effects at play. One such effect is Pauli blocking, a suppression of neutrino-nucleon interactions due to the Pauli exclusion principle. Modelling the nucleus as an ideal Fermi gas of protons and neutrons, one can take Pauli blocking effects into account by requiring that the hit nucleon cannot be in a state which is already occupied \cite{Brown:1971qr}. This requirement is implemented in our calculations by a simple replacement of the differential diffractive cross section
\begin{align*}
\frac{\dd^2 \sigma_{\nu  \mathrm{d}}}{\dd Q^2 \dd \hat{s}}\to f (|\vec{q}|) \, \frac{\dd^2 \sigma_{\nu  \mathrm{d}}}{\dd Q^2 \dd \hat{s}},
\end{align*}
where $|\vec{q}|$ is the magnitude of the transferred three-momentum in the lab frame. In particular, following \cite{Brown:1971qr}, assuming an equal density of neutrons and protons, we have
\begin{equation}
f (|\vec{q}|) = \begin{cases} \displaystyle
                    \frac{3}{2} \frac{|\vec{q}|}{2 \, k_F} - \frac{1}{2} \left( \frac{|\vec{q}|}{2 \, k_F} \right)^3 ,\, &\mathrm{if }\;\; |\vec{q}| < 2\, k_F\, ,\\
                    1,\, &\mathrm{if }\;\; |\vec{q}| \geq 2 \, k_F\, ,
                \end{cases}
\end{equation}
where $k_F$ is the Fermi momentum of the gas, taken to be $235$ MeV. This is a rather low value of $k_F$ and the assumption of equal density of neutrons and protons must be taken with care for heavy nuclei. We refrain from trying to model any additional nuclear effects as we believe that this is the dominant effect on the total diffractive rate, particularly when requiring no hadronic activity in the event. The net result is a reduction of the diffractive cross section by about $50\%$ for protons and $20\%$ for neutrons. 

\begin{figure}[t]
\centering\includegraphics[width=\textwidth]{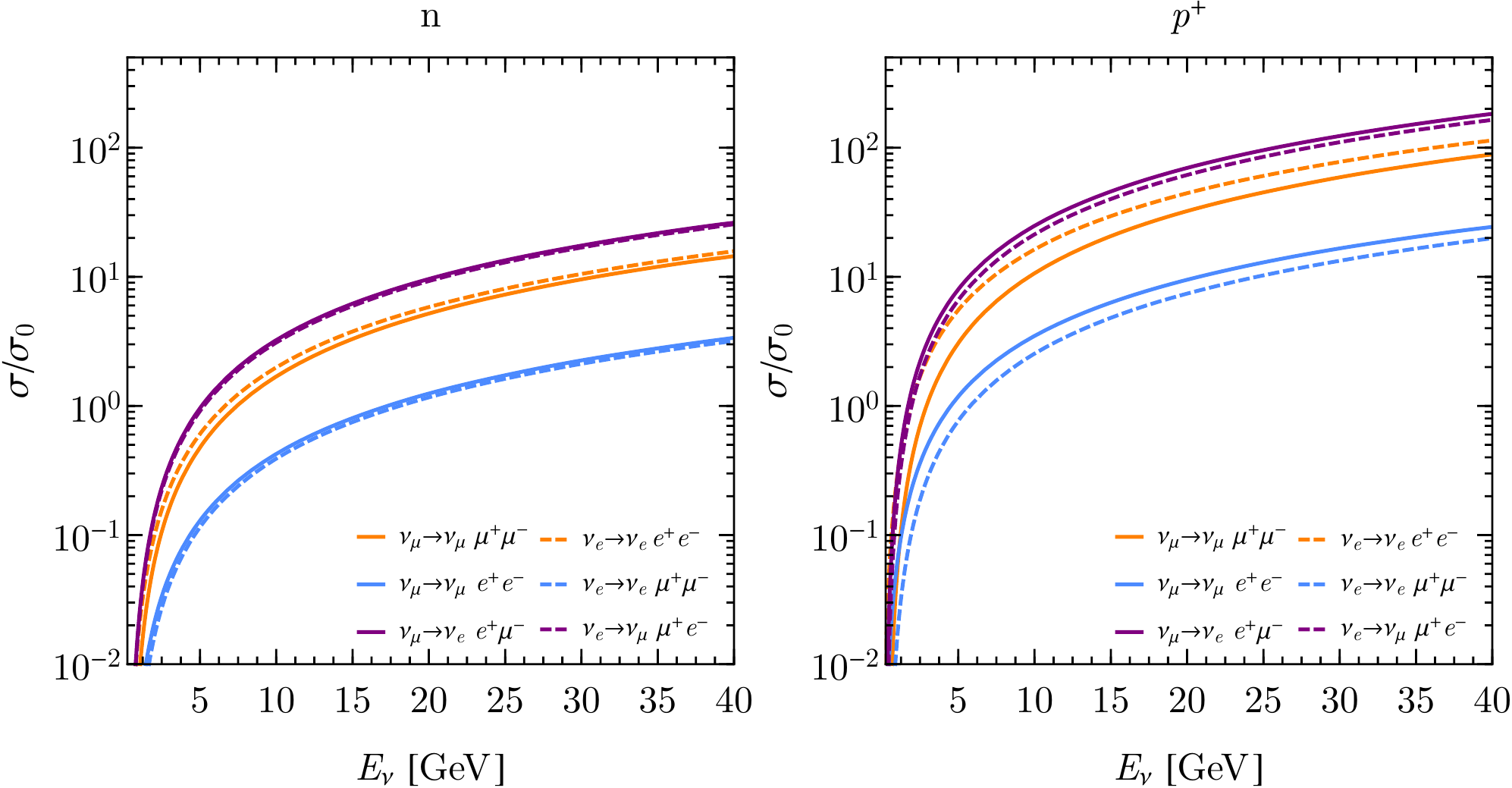}
\caption{Cross sections for diffractive neutrino trident production on neutrons (left) and protons (right), including Pauli blocking effects as described in the text, normalized to $\sigma_0 =  10^{-44}$ cm$^2$. The full (dashed) lines correspond to the scattering of an incoming $\nu_\mu$ ($\nu_e$) produced by the NC (light-blue), CC (purple), and CC+NC (orange) SM interactions. \label{fig:dif_xsec}}
\end{figure}

Our final cross sections for this regime can be seen in \reffig{fig:dif_xsec}. One can clearly see that the neutron contribution is subdominant, and that, up to factors of $Z^2$, the proton one is comparable to the coherent cross section. Note that now the typical values of $Q^2$ are much larger than in the coherent regime and the impact of the final-state lepton masses is much smaller. 

\subsection{Breakdown of the EPA \label{sec:EPAbreakdown}}

In order to understand the breakdown of the EPA in the neutrino trident case, let us first remind briefly the reader about the Weizs\"acker--Williams method of equivalent photons in Quantum Electrodynamics (QED)~\cite{vonWeizsacker:1934nji,Williams:1934ad}, and the main reason for its validity in that theory. The EPA, first introduced by E.\ Fermi~\cite{Fermi:1924tc}, is based on a simple principle: when an ultra-relativistic particle $P_i$ approaches a charged system $C_s$, like a nucleus, it will perceive the electromagnetic fields as nearly transverse, similar to the fields of a pulse of radiation, {\it i.e.},  as an on-shell photon. Therefore, it is possible to obtain an approximate total cross section for the inelastic scattering process producing a set of final particles $P_f$, $\sigma_{\rm t}(P_i + C_s \to P_f + C_s)$, by computing the scattering of the incoming particle with a real photon integrated over the energy spectrum of the off-shell photons,
\begin{align}
	\sigma_{\rm t}(P_i + C_s \to P_f + C_s)\approx\int\, dP(Q^2,\hat{s})\,\sigma_\gamma(P_i + \gamma \to P_f; \hat{s}, Q^2 = 0),
\end{align}
where the photo-production cross section for the process $P_i + \gamma \to P_f$, 
$\sigma_\gamma(P_i + \gamma \to P_f; \hat{s}, Q^2 = 0)$, 
depends on the center-of-mass energy of the $P_i$--photon system, $\sqrt{\hat{s}}$. Here $dP(Q^2,\hat{s})$ corresponds to the energy spectrum of the virtual photons, that is, the probability of emission of a virtual photon with transferred four-momentum $Q^2$ resulting in an center-of-mass energy $\sqrt{\hat{s}}$.
For trident scattering off a nuclear target, this probability can be approximated by~\cite{Belusevic:1987cw,Altmannshofer:2014pba}
\begin{align}\label{eq:GenEPA}
	dP(Q^2,\hat{s})=\frac{Z^2e^2}{4\pi^2}|F (Q^2)|^2\,\frac{d\hat{s}}{\hat{s}}\,\frac{dQ^2}{Q^2}\, .
\end{align}
A crucial fact in QED is that the cross section $\sigma_\gamma^{\rm QED}(P_i + \gamma \to P_f; \hat{s},0)$ is inversely proportional to $\hat{s}$,
\begin{align*}
	\sigma_\gamma^{\rm QED}(P_i + \gamma \to P_f; \hat{s},0) \propto \frac{1}{\hat{s}}\,.
\end{align*}
We see clearly that small values of $\hat{s}$ and consequently of the transferred four-momentum $Q^2$ dominate the cross section. Hence, the on-shell contribution is much more significant 
than the off-shell one, so the EPA will be valid and give the correct cross  section 
estimate for any QED process. 

Now, let us consider the case of neutrino trident production. In this case, the equivalent-photon cross section in the four-point interaction limit has a completely opposite dependence on the center-of-mass energy; it is \emph{proportional} to $\hat{s}$,
\begin{align*}
	\sigma_\gamma^{\rm FL}(P_i + \gamma \to P_f; \hat{s}, 0)\propto G_{\rm F}^2\, \hat{s}\, .
\end{align*}
This dependence is a manifestation of the unitarity violation in the Fermi theory. Therefore, we can see that for weak processes larger values of $\hat{s}$, and, consequently, larger values of $Q^2$ are more significant \cite{Kozhushner:1962aa, Shabalin:1963aa}.  The EPA is then generally not valid for the neutrino trident production, as the virtual photon contribution dominates over the real one. Nevertheless, one may wonder if there is a situation  in which the EPA can give a reasonable estimate for a neutrino trident process. 
As noticed in the early literature \cite{Kozhushner:1962aa, Shabalin:1963aa}, the presence of the nuclear form factor introduces a cut in the transferred momentum which, in turn, makes the EPA applicable for the specific case of the dimuon channel in the coherent regime. Let us discuss this in more detail. 

Recalling our exact decomposition, \refeq{eq:full_diff_xsec}, it is necessary to consider two assumptions for implementing the EPA \cite{Kozhushner:1962aa}:
\begin{enumerate} 
\item The longitudinal polarization contribution to the cross section can be neglected, i.e., $\sigma_{\nu\gamma}^\mathrm{L}(Q^2,\hat{s})\approx 0$;
\item The transverse polarization contribution to the cross section can be taken to be on-shell, i.e., $\sigma^\text{T}_{\nu\gamma}(Q^2,\hat{s}) \approx \sigma^\text{T}_{\nu\gamma}(0,\hat{s})$. 
\end{enumerate}
Assuming for now that these approximations hold, we can find a simplified expression for the coherent neutrino-target process, described by Eqs.~(\ref{eq:full_diff_xsec}) and (\ref{eq:hcoh}), in terms of the photon-neutrino cross section\footnote{An analogous expression can be obtained for the diffractive regime from Eq.~(\ref{eq:dcoh}).}:
\begin{align}     
\sigma_\text{EPA} = \frac{Z^2e^2}{4\pi^2}\int_{m_L^2}^{\hat{s}_{\rm max}} \frac{d\hat{s}}{\hat{s}}\,
\sigma^\mathrm{T}_{\nu\gamma}(0,\hat{s})
\int_{(\hat{s}/2E_\nu)^2}^{Q^2_{\rm max}}\frac{|F (Q^2)|^2}{Q^4} \left[ Q^2(1-y) - M_{\cal H}^2y^2\right]dQ^2\, , 
\end{align}
where we introduced the fractional change of the nucleus energy $y$, defined as $\hat{s} = (s-M_{\cal H}^2)y$, and the integration limits can be obtained from \eqref{eq:qslimts} after considering that $m_L^2\ll E_\nu M_{\cal H}$. Keeping only the leading terms in the small parameter $y$ \cite{Belusevic:1987cw}, we recover the EPA applied to the neutrino trident case
\begin{align} \label{eq:EPA_bad}
\sigma_\text{EPA} = \int \sigma^\mathrm{T}_{\nu\gamma}(0,\hat{s}) \, dP(Q^2,\hat{s})\, ,
\end{align}
where $dP(Q^2,\hat{s})$ is given in Eq.~(\ref{eq:GenEPA}). The EPA in the form of \refeq{eq:EPA_bad} has been used in trident calculations for the coherent dimuon channel \cite{Altmannshofer:2014pba} as well as for coherent mixed- and electron-flavour trident modes and diffractive trident modes \cite{Magill:2016hgc}.  Using our decomposition, we can explicitly compute both $\sigma^\mathrm{L}_{\nu \gamma}$ and $\sigma^\mathrm{T}_{\nu \gamma}$ and verify if the EPA conditions are satisfied for any channel and, if they are not, quantify the error introduced by making this approximation. For that purpose, we will compare the results of the full calculation, \refeq{eq:full_diff_xsec}, with the EPA results, \refeq{eq:EPA_bad}, by computing the following ratios in the physical region of the $(Q,\hat{s})$ plane,
\begin{align}\label{eq:ratios}
		\frac{\sigma^{\rm L}(Q^2,\hat{s})\,h_{\rm c}^{\rm L}(Q^2,\hat{s})}{\sigma^{\rm T}(Q^2,\hat{s})\,h_{\rm c}^{\rm T}(Q^2,\hat{s})}\, , \quad \frac{\sigma^\mathrm{T}_{\nu\gamma}(Q^2,\hat{s})}{\sigma^\mathrm{T}_{\nu\gamma}(0,\hat{s})}\, .
\end{align} 
The first ratio in Eq.\ \eqref{eq:ratios} will indicate where the longitudinal contribution can be neglected compared to the transverse one; while, the second ratio will show where the transverse contribution behaves as an on-shell photon. 

\begin{figure}[t]
\centering
\includegraphics[width=\textwidth]{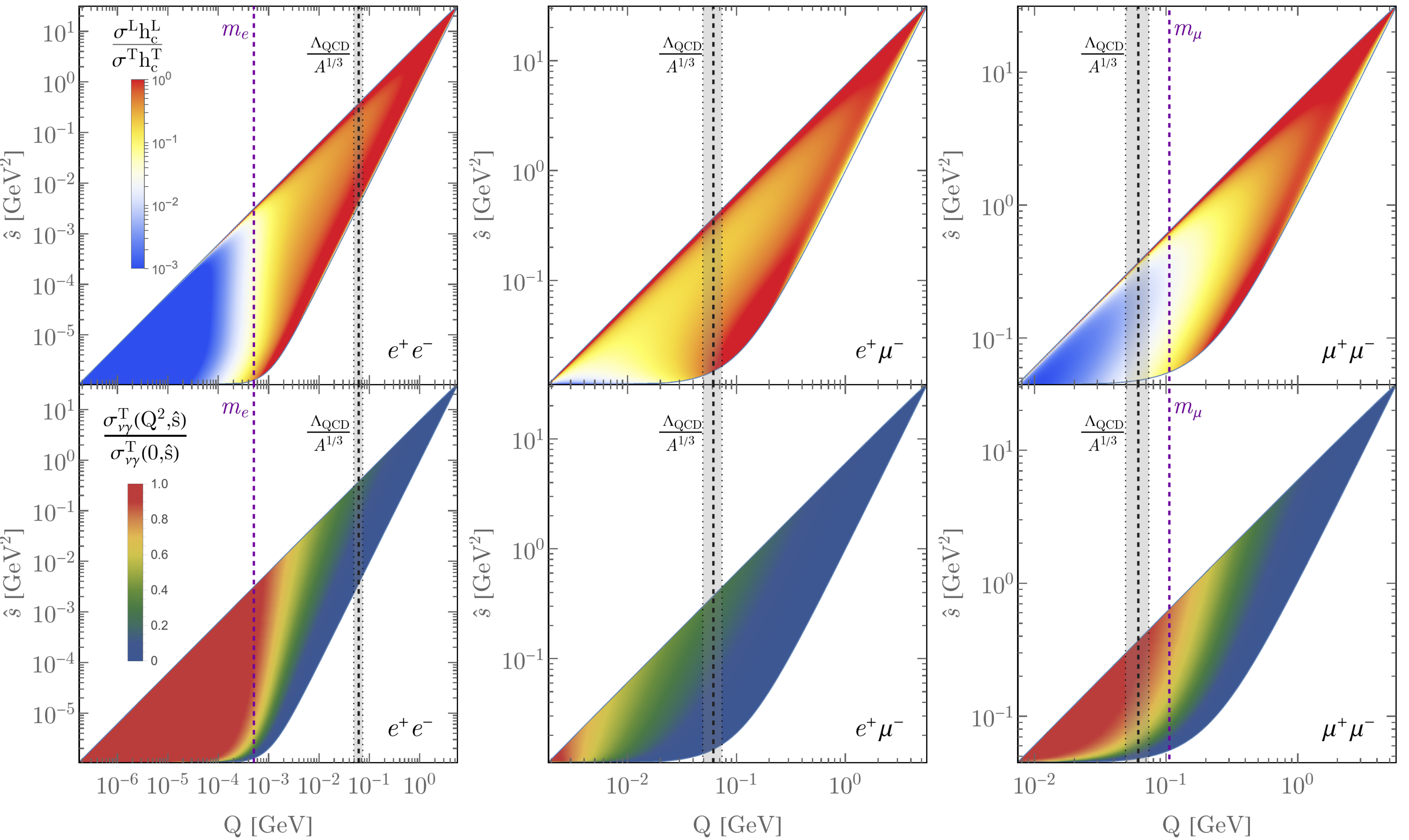}%
\caption{\label{fig:4PSvsEPA} Comparison between the full calculation of the trident production 
coherent cross section and the EPA in the kinematically allowed region of the $(Q,\hat{s})$ plane for an incoming $\nu_\mu$ with fixed energy $E_\nu=3$ GeV colliding with an $^{40}$Ar target. 
The left, middle and right panels correspond to the dielectron, mixed and dimuon final-states, respectively. The top panels correspond to the comparison between the longitudinal and transverse contributions while the bottom ones show the ratio between the transverse cross sections computed for an specific value of $Q$ with the cross section for an on-shell photon. The thick black dashed lines correspond to the cut in the $Q^2$ integration at $\Lambda_{\rm QCD}^2/ A^{2/3}$, and the shadowed region around these lines account for a variation of $20\%$ in the value of this cut. The purple dashed lines are for $Q=m_\alpha$, $\alpha=e,\mu$ for the unmixed cases.}
\end{figure}

As an illustration of the general behaviour, we show in Fig.\ \ref{fig:4PSvsEPA} those ratios 
of cross sections for an incoming $\nu_\mu$ of fixed energy $E_\nu=3$ GeV colliding coherently with an $^{40}$Ar target, for the dielectron (left panels), mixed  (middle panels) and dimuon  (right panels)
channels. On the top panels of Fig.\ \ref{fig:4PSvsEPA} we see that the longitudinal component can be neglected for $Q\lesssim m_\alpha$, for the dielectron and dimuon channels, $\alpha=e,\mu$, while in the mixed case there is a much less pronounced hierarchy between the transverse and longitudinal components. On the bottom panels we have the comparison between on-shell and off-shell transverse photo-production cross sections. Again, we find that the EPA is only valid for $Q \lesssim m_\alpha$ for the dielectron and dimuon channels. For the mixed case, there is only a very small region in $Q < 10^{-2}$\,GeV for which the off-shell transverse cross section is comparable to the on-shell one. This relative suppression of the off-shell cross section can be understood by noticing that $Q$ enters the lepton propagators, suppressing the process for $Q \gtrsim m_\alpha$. For mixed channels it is then the smallest mass scale ($m_e$) that dictates the fall-off of the matrix element in $Q$, whilst the heaviest mass ($m_\mu$) defines the phase space boundaries, rendering most of this phase space incompatible with the EPA assumptions.   

These results explicitly show that the EPA is, in principle, not suitable for any neutrino trident process as it can overestimate the cross section quite substantially by treating the photo-production cross section at large $Q^2$ as on-shell. However, as previously mentioned, in the coherent regime the nuclear form factor introduces a strong suppression for large values of $Q^2$. In general, this dominates the behaviour of the cross sections for values of $Q^2$ smaller than the purely kinematic limit, $Q^2_{\rm max}$, and of the order of $\Lambda_{\rm QCD}/ A^{1/3}\approx 0.06$ GeV for coherent scattering on $^{40}$Ar. In the dimuon case, the latter scale happens to be smaller than the charged lepton masses, implying that the region where the EPA breaks down is heavily suppressed due to the nuclear form factor. The same cannot be said about coherent trident channels involving electrons, as the nuclear form factor suppression happens for much larger values of $Q$ than the EPA breakdown. Furthermore, for diffractive scattering the nucleon form factors suppress the cross sections only for much larger $Q$ values, $Q\approx 0.8$ GeV. The effective range of integration then includes a significant region where the EPA assumptions are invalid, leading to an overestimation of the diffractive cross section for every process regardless of the flavours of their final-state charged leptons. 

\begin{figure}[t]
	\centering
	\includegraphics[width=\textwidth]{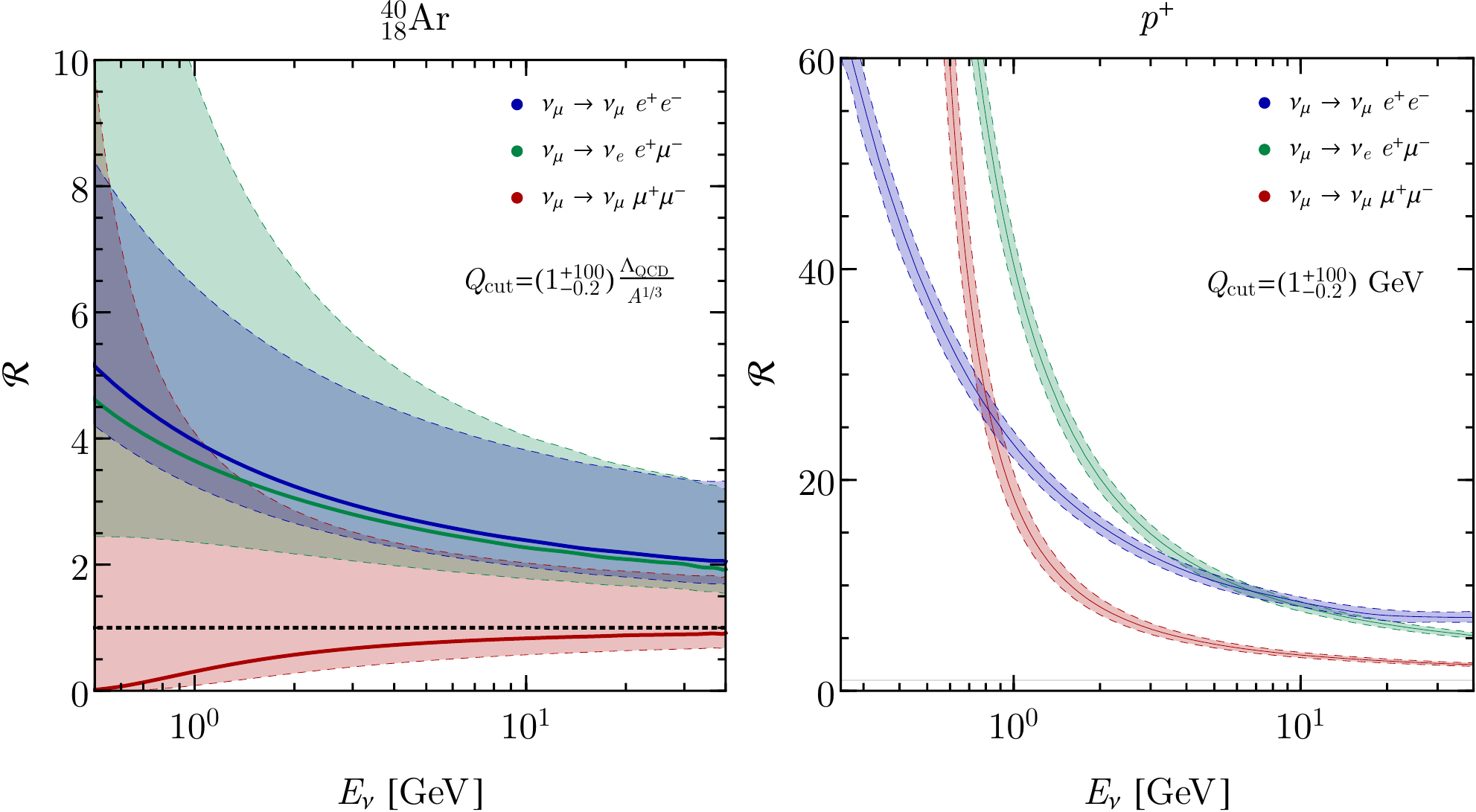}%
	\caption{\label{fig:comparison4PS_EPA} 
    Ratio $\mathcal{R}$ of the trident cross section calculated using 
    the EPA to the full four-body calculation. 
    Left panel: Ratio in the coherent regime on $^{40}$Ar. The full curves correspond to the central value of $Q_{\rm cut}$, and the upper (lower) boundary corresponds to a choice 100 times larger ($20\%$ smaller). 
Right  panel: Ratio in the diffractive regime for scattering on protons, where the full curves corresponds to the central value of $1.0$ GeV, and the upper (lower) boundary corresponds to a choice 100 times larger ($20\%$ smaller); we have taken the lower limit in the integration on $Q$ to match the choice of the coherent regime. A guide to the eye at $\mathcal{R} = 1$ is also shown.}
\end{figure}

In some calculations, artificial cuts have been imposed on the range of $Q^2$, affecting the validity of the EPA. In Ref. \cite{Magill:2016hgc}, it is claimed that to avoid double counting between different regimes, an artificial cut must be imposed, lowering the upper limit of integration in $Q^2$. Ref.~\cite{Magill:2016hgc} chooses a value of $Q^{\rm cut}_{\rm max} = \Lambda_{\rm QCD}/ A^{1/3}$ in the coherent regime (black thick dashed lines in Fig.\ \ref{fig:4PSvsEPA}), and $Q^{\rm cut}_{\rm min}= {\rm max}\left( \Lambda_{\rm QCD}/ A^{1/3}, \hat{s}/2E_\nu\right)$ and $Q^{\rm cut}_{\rm max} = 1.0$ GeV in the diffractive regime. We believe that no such cut is required on physical grounds\footnote{It should be noted that the coherent and diffractive regimes have different phase space boundaries and that the form factors should guarantee their independence.}, and their presence will impact the EPA cross section quite dramatically. Let us first consider the dimuon case in the coherent regime, where the EPA assumptions hold reasonably well in the relevant parts of phase space. By introducing a value for $Q^{\rm cut}_{\rm max}$ we would be decreasing the total relevant phase space for the process, reducing the total cross section. Therefore, despite the EPA tendency to overestimate the cross section in this channel, an artificial cut in $Q^2$ can actually lead to an underestimation of the cross section. In the electron channels, where the EPA breakdown is much more dramatic, we can expect that the overestimation of the cross section by the EPA is reduced by the cut $Q^{\rm cut}_{\rm max}$. In fact, one way to improve the EPA for the dielectron channel is to artificially cut on the $Q^2$ integral around the region where the ap\-pro\-xi\-ma\-tion breaks down \cite{Frixione:1993yw}. This cut does then improve the coherent EPA calculation by decreasing the overestimation of the cross section. However, an energy independent cut cannot provide a good estimate of the cross section over all values of $E_\nu$. To illustrate our point and to quantify the errors induced by the EPA, we show on the left panel of \reffig{fig:comparison4PS_EPA} the ratio $\mathcal{R}$ of the trident cross section calculated using the EPA with an artificial cut at $Q^2_\text{cut}$, as performed in \cite{Magill:2016hgc}, to the full calculation used in this work as a function of the incoming neutrino energy:
 \begin{equation}
 \mathcal{R} = \frac{\sigma_{\rm EPA} (E_\nu) \vert_{Q_{\rm cut}}}{\sigma_{\rm 4PS} (E_\nu)}\,.
 \end{equation}
 In this plot we vary the artificial cut on $Q^2$ around the choice of \cite{Magill:2016hgc} (shown as the central dashed line) in two ways. First we reduce it by $20 \%$, and then increase it by a large factor, recovering the case with no $Q^2$ cut. From this, our conclusions about the validity of the approximation are confirmed, and it becomes evident that the trident coherent cross section is very sensitive to the choice of $Q^2_{\mathrm{cut}}$. In particular, the EPA with all the assumptions that lead to \refeq{eq:EPA_bad} and the absence of a $Q^2$ cut can lead to an overestimation of all trident channels, including the dimuon one. Once the cut is implemented, however, the approximation becomes better for the dimuon channel, but still unacceptable for the electron ones. It is also clear that an energy independent cut cannot give the correct cross section at all energies. This is particularly troublesome for detectors subjected to a neutrino flux covering a wide energy range such as the near detectors for DUNE and  MINOS or MINER$\nu$A. Moreover, \refeq{eq:EPA_bad} fails at low energies, and generally, overestimates the coherent cross sections by at least  200\%. At these energies, one must be wary of the additional approximations in \refeq{eq:EPA_bad} regarding the integration limits and the small $y$ limit.     

On the right panel of \reffig{fig:comparison4PS_EPA} we illustrate what happens in the diffractive regime, where the nucleon form factors impact the cross section at much larger values of $Q^2$ and have a slower fall-off. We see that the diffractive cross section is dramatically overestimated over the full range of $E_\nu$ considered and for any trident mode. The discrepancy is particularly important for $E_\nu \lesssim$ 5 GeV and larger than in the coherent regime by at least an order of magnitude\footnote{There are some differences in the treatment of the hadronic system between the EPA calculation in \cite{Magill:2016hgc} and the one presented here. However, these differences are of the order 10\% to 20\%. Note also that we do not implement any Pauli blocking when calculating $\mathcal{R}$ to avoid ambiguities over the choice of the range of $Q^2$.}. We also see that the cuts on $Q^2$ impact the EPA calculation much less dramatically, and that its use is unlikely to yield the correct result.

Given these problems with both coherent and diffractive cross section calculations due to the breakdown of the EPA for trident production, in what follows we will use the complete four-body calculation.

\subsection{Coherent versus Diffractive Scattering in Trident Production}
\label{subsec:cohdiff}

\begin{figure}[t]
	\centering
	\includegraphics[width=\textwidth]{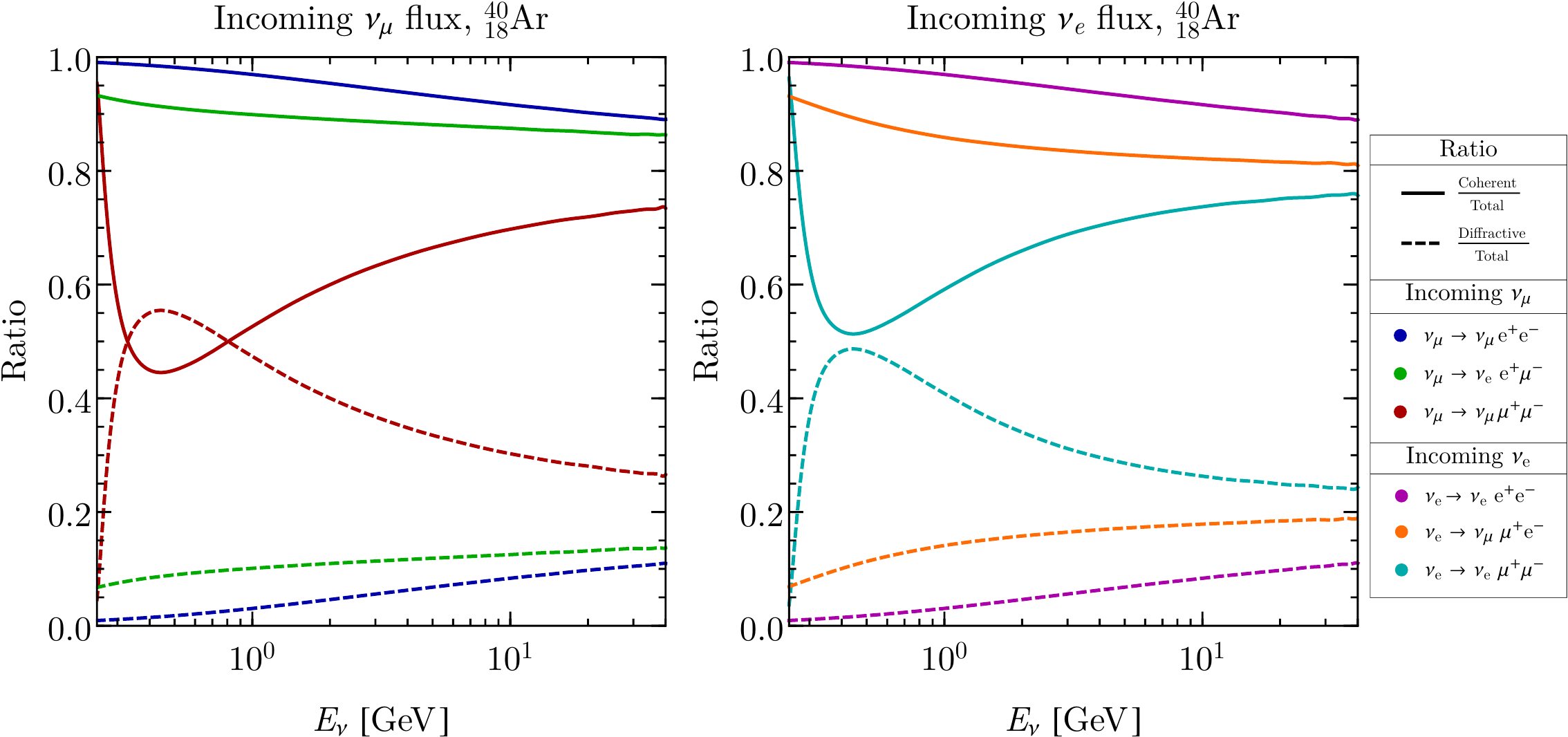}%
	\caption{\label{fig:RatioCDvsT} On the left (right) panel we show the
    ratio of the coherent (full lines) and the diffractive (dashed lines) contributions to the total trident cross section for an incoming flux of $\nu_\mu$($\nu_e$) as a function of $E_\nu$ for an 
    $^{40}$Ar target.}
\end{figure}

Let us  now comment on the significance of  the coherent and diffractive contributions to the total cross for the different trident channels. 
In Fig.\ \ref{fig:RatioCDvsT} we present the ratio of the coherent and the diffractive scattering cross sections to the total cross section for an $^{40}$Ar target for an incoming $\nu_\mu$ (left) and $\nu_e$ (right)  neutrino. We can see that the coherent regime dominates at all neutrino energies when there is an electron in the final-state, especially in the dielectron case. 
This can be explained by noting that the $Q^2$ necessary to create an electron pair is smaller than the one needed to create a muon; thus, coherent scattering is more likely to occur for this mode. Conversely, 
as one needs larger momentum transferred to produce a muon (either accompanied by an electron 
or another muon) the  diffractive regime becomes more likely in these modes, as we can explicitly 
see in Fig.\ \ref{fig:RatioCDvsT}. 
Because of this effect the diffractive contribution is $\lesssim$ 10\%, except for the 
dimuon channel where it can be between $30$ and $40$\% in most of the energy region.
Furthermore, when we compare the two incoming types of neutrinos, we see that for an incoming $\nu_\mu$ the diffractive contribution is larger than the coherent one in the range $0.3\ {\rm GeV}\lesssim E_\nu \lesssim 0.8$ GeV, while for an incoming $\nu_e$ this never happens. 
This difference can be explained by the fact that CC and NC contributions  are simultaneously present for the scattering of an initial $\nu_\mu$ creating a muon pair, whereas
for an initial $\nu_e$ creating a muon pair, we will only have the NC contribution, see Table \ref{tab:tridentmodes}.

An important difference between the coherent and diffractive regimes will be in their hadronic signatures in the detector. Neutrino trident production is usually associated with zero hadronic energy at the vertex, a feature that proved very useful in reducing backgrounds in previous measurements. Whilst this is a natural assumption for the coherent regime, it need not be the case in the diffractive one. In fact, in the latter it is likely that the struck nucleon is ejected from the nucleus in a significant fraction of events with $Q$ exceeding the nuclear binding energy
\footnote{The peak of our diffractive $Q^2$ distributions happens at around $Q \approx 300$ MeV, much beyond the typical binding energy for Ar (see \refapp{app:distributions}). Without Pauli suppression, however, we expect this value to drop.}. Since the dominant diffractive contribution comes from scattering on protons, these could then be visible in the detector if their energies are above threshold. On the other hand, the struck nucleon is subject to many nuclear effects which may significantly affect the hadronic signature, such as interactions of the struck nucleon in the nuclear medium as well as reabsorption. Our calculation of Pauli blocking, for example, shows large suppressions ($\sim 50\%$) precisely in the low $Q^2$ region, usually associated with no hadronic activity. This then raises the question of how well one can predict the hadronic signatures of diffractive events given the difficulty in modelling the nuclear environment. We therefore do not commit to an estimate of the number of diffractive events that would have a coherent-like hadronic signature, but merely point out that this might introduce additional uncertainties in the calculation, especially in the $\mu^+ \mu^-$ channel where the diffractive contribution is comparable to the coherent one. Finally, from now on we will refer to the number of trident events with no hadronic activity as coherent-like, where this number can range from coherent only to coherent plus all diffractive events.

\section{Trident Events in LAr Detectors}
\label{sec:LAr}

\begin{figure}[t]
\centering
\includegraphics[width=1.\textwidth]{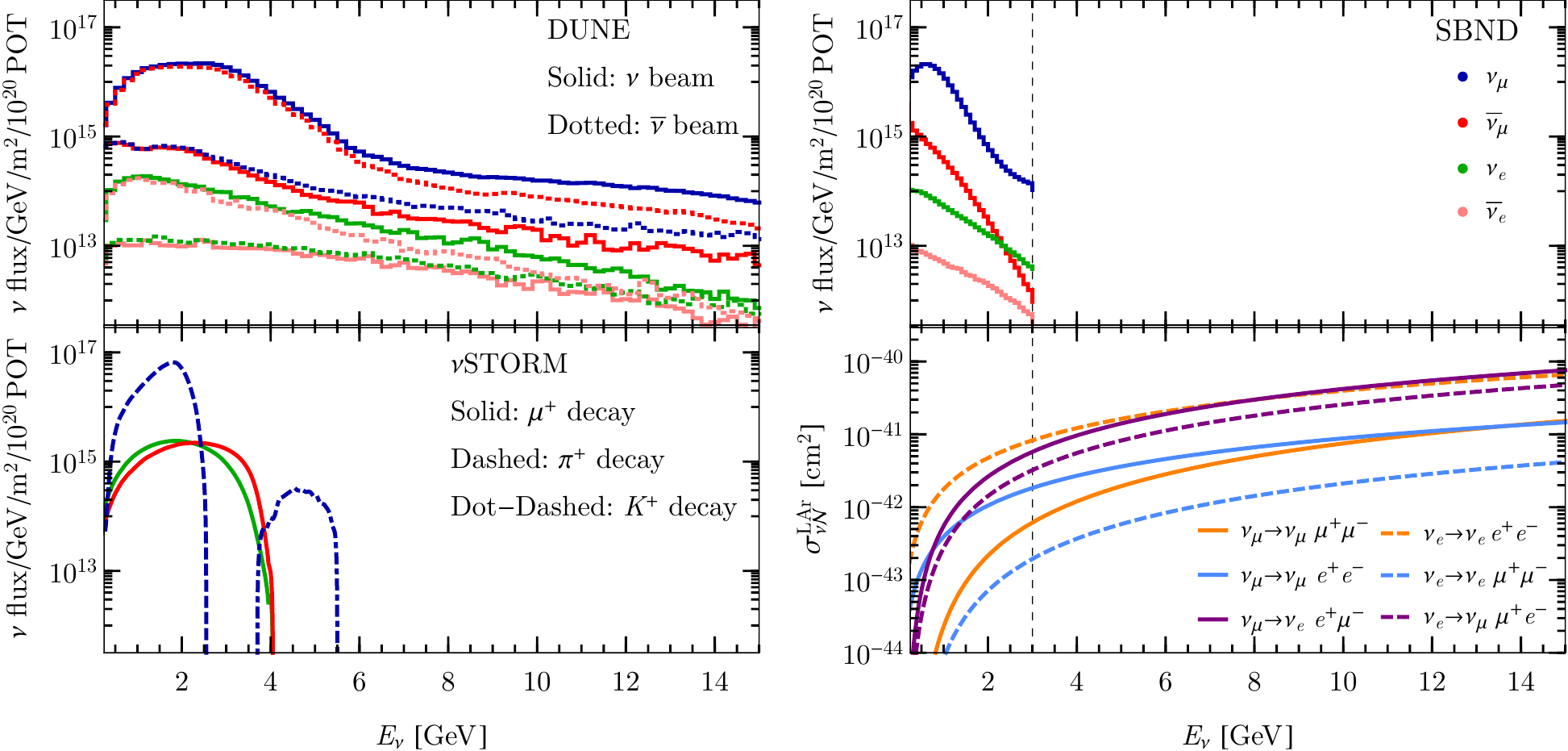}
\caption{\label{fig:LAr}Energy distribution of the neutrino fluxes at the position of 
the LAr detectors DUNE (top left, \cite{DUNE:flux}), SBND (top right,\cite{SBNproposal}) and $\nu$STORM (bottom left, \cite{nuSTORM2017}) and of the cross sections for the various trident modes (bottom right). The fluxes at $\mu$BooNE and ICARUS are similar to the one shown for SBND when normalized over distance.}
\end{figure}

In this section we calculate the total number of expected  trident events for some present and future LAr detectors with different fiducial masses, total exposures and beamlines. In Table~\ref{tab:LAr} we specify the values used for each set-up and in Fig.~\ref{fig:LAr} we show the total production cross section for each neutrino trident mode of Table~\ref{tab:tridentmodes}
as well as the neutrino fluxes as a function of $E_\nu$ at the position of each experiment.  

\subsection{Event Rates}
\label{subsec:rates}

The total number of trident events, $N^{\text{\Neptune}}_{\rm X}$, expected for a given trident mode at any detector is written as  
\begin{eqnarray}
\label{eq:nevents}
N^{\text{\Neptune}}_{\rm X}={\rm Norm}\times\int dE_\nu \, \sigma_{\nu {\rm X}}(E_\nu) \frac{d\phi_{\nu}(E_\nu)}{dE_\nu}\epsilon(E_\nu)\,,
\end{eqnarray}
where $\sigma_{\nu {\rm X}}$ can be the trident total (${\rm X}={\cal N}$), coherent ($\mathrm{X=c}$) or diffractive ($\mathrm{X=d}$) cross sections 
for a given mode, $\phi_{\nu}$ is the flux of the incoming neutrino and $\epsilon(E_\nu)$ is the efficiency of detection of the charged leptons. In the calculations of this section, we assume an efficiency of $100\%$\footnote{See \refsec{subsec:kine} for a discussion on the detection efficiencies for trident events and backgrounds.}.
The normalization is calculated as 
$${\rm Norm}= {\rm{Exposure}}~[{\rm{POT}}] \times \frac{{\rm{Fiducial~Detector~Mass}\times N_A}}{m_{\rm T}} \left[{\rm{target~particles}}\right],$$
where $m_{\rm T}$ is the molar mass of the target particle and $N_A$ is Avogadro's number.
Two features of the cross sections are important for the event rate calculation: 
threshold effects, especially for channels involving muons in the final-state,
and cross section's growth with energy. In particular, we expect higher trident event rates for experiments with higher energy neutrino beams. 

We start our study with the three detectors of the SBN program, one of which, $\mu$BooNE, is already installed and taking data at Fermilab. These three LAr time projection chamber detectors are located along the Booster Neutrino Beam line which is by now a well-understood source, having the focus of active research for over 15 years. 
Although the number of trident events expected in these detectors is rather low, they may offer one of the first opportunities to study trident events in LAr, as well as to better understand their backgrounds in this medium and to devise improved analysis techniques.
After that we study the proposed near detector for DUNE. This turns out to be the most important LAr detector for trident production since it will provide the highest number of events in both neutrino and antineutrino modes. 
Finally, having in mind the novel flavour composition of neutrino beams from muon facilities, we investigate trident rates at a 100~t LAr detector for the $\nu$STORM project. This last facility could offer a very well understood neutrino beam with as many electron neutrinos as muon antineutrinos from muon decays, creating new possibilities for trident scattering measurements.
\begin{table}[t]
\begin{center}
\scalebox{0.9}{
\begin{tabular}{|c|c|c|c|c|}
\hline\hline
\bf Experiment& \bf Baseline (m) & \bf Total Exposure (POT) & \bf Fiducial Mass (t) & \bf $\mathbf{E_\nu}$ (GeV)\\\hline\hline
SBND & 110 & $6.6\times 10^{20}$ & 112 & $0-3$\\\hline
$\mu$BooNE & 470 & $1.32\times 10^{21}$ & 89& $0-3$\\\hline
ICARUS & 600 & $6.6\times 10^{20}$ & 476 & $0-3$\\\hline
DUNE & 574  & $12.81~(12.81)\times 10^{21}$& 50 & $0-40$ \\\hline
$\nu$STORM & 50  & $10^{21}$ & 100& $0-6$\\\hline\hline
\end{tabular}}
\end{center}
\caption{\label{tab:LAr} Summary of the LAr detectors set-up and values assumed in our calculations.
The POT numbers are given for a neutrino (antineutrino) beam.}
\end{table}

\subsubsection{The SBN Program}
\label{subsubsec:SBND}

The SBN Program at Fermilab is a joint endeavour by three collaborations ICARUS, $\mu$BooNE and 
SBND to perform searches for eV-sterile neutrinos and study neutrino-Ar cross sections \cite{SBNproposal}. As can be seen in Tab.~\ref{tab:LAr}, SBND has the shortest baseline (110 m) and therefore the largest neutrino fluxes (shown in Fig.~\ref{fig:LAr} and taken from Fig. 3 of \cite{SBNproposal}). The largest detector, ICARUS, is also the one with the longest baseline (600 m) and consequently subject to the lowest neutrino fluxes.
The ratio between the fluxes at the different detectors are  $\phi_{\mu\rm{BooNE}}/\phi_{\rm{SBND}}=5$\% and $\phi_{\rm{ICARUS}}/\phi_{\rm{SBND}}=3$\%.
The neutrino beam composition is about 93\% of $\nu_\mu$,  6\% of $\overline\nu_\mu$ and  
$1\%$ of $\nu_e+\overline{\nu}_e$. 

Considering the difference in fluxes and the total number of targets in each of these 
detectors, one can estimate the following ratios of trident events: 
${N^\text{\Neptune}_{\mu\rm{BooNE}}}/{N^\text{\Neptune}_{\rm{SBND}}}\sim 8$\% and ${N^\text{\Neptune}_{\rm{ICARUS}}}/{N^\text{\Neptune}_{\rm{SBND}}}\sim 10$\%. Unfortunately, 
since the fluxes are peaked at a rather low energy ($E_\nu \lesssim 1$ GeV), where the trident  
cross sections are still quite small ($\lesssim 10^{-42}$ cm$^2$) we expect very few 
trident events produced.
The exact number of trident events for those detectors according to our calculations is 
presented in Tab.~\ref{tab:LArrates}. For each trident channel the first (second) row
shows the number of coherent (diffractive) events. As expected, less than a total 
of 20 events across all channels can be detected by SBND, and a negligible rate of events is expected at $\mu$BooNE and ICARUS. 

\subsubsection{DUNE Near Detector}
\label{subsubsec:DUNE}

\begin{table}[t]
\begin{center}
\scalebox{0.9}{
\begin{tabular}{|cccccc|}
\hline\hline
		\bf Channel & \bf SBND& \bf $\mu$BooNE & \bf ICARUS & \bf DUNE ND &\bf  $\nu$STORM ND \\ \hline \hline
		$\nu_\mu\to\nu_e e^+ \mu^-$& $10$ &$0.7$ &$1$ &$2844 ~ (235)$ & $159$ \\
        &$2$ &$0.1$ &$0.2$ &$654 ~ (56)$& $35$\\\hline
        $\overline\nu_\mu\to\overline\nu_e e^- \mu^+$&$0.4$ &$0.02$ &$0.04$ &$122~(2051)$ & $23$\\
        &$0.08$ &$0.005$ &$0.008$ &$29~(468)$ & $5$\\\hline
	$\nu_e\to\nu_\mu e^- \mu^+$& $0.05$ &$0.003$ &$0.004$ &$22~(7)$ & $9$\\
        &$0.01$ &$0.0008$ &$0.001$ &$7~(2)$ & $3$\\\hline
	$\overline\nu_e\to\overline\nu_\mu e^+ \mu^-$& $0.005$ &$0.0003$ &$0.0005$ &$5~(14)$ & $-$\\
    &$0.001$ &$0.0001$ &$0.0001$ &$2~(4)$ & $-$\\\hline
    \hline\hline
    {$\rm{Total} \ e^\pm \mu^\mp$}& $10$ &$0.7$ &$1$ &$2993~(2307)$ & $191$ \\
    &$2$ &$0.1$ &$0.2$ &$692~(530)$ & $41$\\\hline
    \hline
		$\nu_\mu\to\nu_\mu e^+ e^-$& $6$ &$0.4$ &$0.7$ &$913~(58)$ & $73$ \\
        &$0.7$ &$0.04$ &$0.07$ &$128~(9)$ & $9$\\\hline
        $\overline\nu_\mu\to\overline\nu_\mu e^- e^+$& $0.2$ &$0.01$ &$0.02$ &$34~(695)$ & $9$\\
        &$0.03$ &$0.001$ &$0.002$ &$5~(95)$ & $1$\\\hline
	$\nu_e\to\nu_e e^- e^+$&$0.2$ &$0.01$ &$0.02$ &$50~(13)$ & $32$ \\
    &$0.02$ &$0.001$ &$0.002$ &$8~(2)$ & $4$\\\hline
	$\overline\nu_e\to\overline\nu_e e^+ e^-$&$0.02$ &$0.001$ &$0.002$ &$10~(34)$ & $-$ \\
    &$0.003$ &$0.0001$ &$0.0002$ &$2~(5)$ & $-$\\\hline
    \hline\hline
    ${\rm{Total}}\  e^+ e^-$& $6$ &$0.4$ &$0.7$ &$1007~(800)$ & $114$\\
    &$0.7$ &$0.0$ &$0.1$ &$143~(111)$ & $14$\\\hline
    \hline    
		$\nu_\mu\to\nu_\mu \mu^+ \mu^-$& $0.4$ &$0.03$ &$0.04$ &$271~(32)$ & $9$ \\
        &$0.4$ &$0.03$ &$0.04$ &$186~(19)$ & $8$\\\hline
        $\overline\nu_\mu\to\overline\nu_\mu \mu^- \mu^+$& $0.01$ &$0.001$ &$0.001$ &$14~(177)$ & $2$\\
        &$0.01$ &$0.0009$ &$0.001$ &$9~(127)$ & $1$\\\hline
 $\nu_e\to\nu_e \mu^+ \mu^-$    &$0.002$ &$0.0001$ &$0.0001$ &$1~(0.5)$ & $0.4$\\  
 &$0.001$ &$0.0001$ &$0.0001$ &$0.7~(0.2)$ & $0.3$\\\hline
        $\overline\nu_e\to\overline\nu_e \mu^+ \mu^-$&$0.0002$ &$0.0000$ &$0.0000$ &$0.3~(0.9)$ & $-$\\
        &$0.0001$ &$0.0000$ &$0.0000$ &$0.2~(0.5)$ & $-$\\\hline
        \hline\hline
    ${\rm{Total}} \ \mu^+ \mu^-$ &$0.4$ &$0.0$ &$0.0$ &$286~(210)$ & $11$ \\
    &$0.4$ &$0.0$ &$0.0$ &$196~(147)$ & $9$\\
    \hline\hline
\end{tabular}}
\end{center}
\caption{\label{tab:LArrates}Total number of \textbf{coherent} (top row) and \textbf{diffractive} (bottom row) trident events expected at different LAr experiments for a given channel.
The numbers in parentheses are for the antineutrino running mode, when present. These calculations  
considered a detector efficiency of 100\%. }
\end{table}

The DUNE experiment will operate with neutrino as well as antineutrino LBNF beams produced by 
directing a 1.2 MW beam of protons onto a fixed target \cite{Acciarri:2016ooe,DUNECDRvolII}. 
The design of the near detector is not finalised, but the current designs favour a mixed technology  detector combining a LAr TPC with a larger tracker module.  In this work, we will assume that DUNE ND is a LAr detector located at $574$ m from the target with a fiducial mass of 50~t \cite{WeberTalk}. As the trident event rate scales with the density of the target, any tracker module will not significantly influence the total event rate, and does not feature in our estimates; although, its presence is assumed to improve reconstruction of final-state muons. Our estimates can be easily scaled for the final design by using \refeq{eq:nevents}.

For the first 6 years of data taking (3 years in the neutrino plus 3 years in the antineutrino 
mode) the collaboration expects $1.83\times 10^{21}$~POT/year with  a plan to upgrade the beam after the 6th year for 2 extra years in each beam mode  with double exposure, making a total of $1.83 \times(3+2\times2)\times 10^{21}~{\rm{POT}}$ for each mode \cite{DUNE:exposure}. We will 
assume the total 10-year exposure in our calculations.
We use the optimized 3-horn fluxes for a beam of 62.4 GeV protons taken from Ref. \cite{DUNE:flux} as the relevant fluxes at the DUNE ND location (see Fig.~\ref{fig:LAr}). The beam composition of the neutrino (antineutrino) beam is about 96\% $\nu_\mu$ ($\overline\nu_\mu$), 4\%  $\overline\nu_\mu$ ($\nu_\mu$) and 1\% $\nu_e+\overline\nu_e$.
 
The number of trident events for DUNE ND can be found in Tab.~\ref{tab:LArrates}. 
The numbers in parentheses correspond to antineutrino beam mode.
Note that although the trident cross sections are the same 
for neutrinos and antineutrinos, the fluxes are a bit lower for the antineutrino beam, as a consequence we predict a lower event rate for this beam\footnote{A similar difference will apply to the processes constituting the background to the trident process, although there is an additional suppression in many channels due to the lower antineutrino cross sections.}.
Due to the much higher energy and wider energy range of the neutrino fluxes at DUNE ND, as compared to the SBN detectors, DUNE can observe a considerable number of trident events, about 300 times the number of trident events expected for SBND just in the neutrino mode. Moreover, the subdominant component of 
each beam mode will also contribute to the signal. For example, we expect to observe $2051$ trident events in the $\overline{\nu}_\mu\to\overline{\nu}_e e^- \mu^+$ channel in the antineutrino mode. However, we also expect 
$235$ events in the $\nu_\mu\to\nu_e e^+ \mu^-$ channel produced by 
the subdominant component of $\nu_\mu$ in the antineutrino beam.
We have considered 100\% detection efficiency here, however, we will see in Sec.~\ref{subsec:bck} that after implementing hadronic vetos, detector thresholds and kinematical cuts to substantially reduce the background we expect an efficiency of about 47\%-65\% on coherent tridents, depending on the channel (see Tab.~\ref{tab:DUNE_ND_NU_BG}).

The mixed flavour trident channel is the one with the highest statistics (more than 6500 events adding 
neutrino and antineutrino beam modes), 18\% of which are produced by diffractive scattering. The dielectron channel comes next with a total of a bit more than 2000 events, 12\%  of which are produced by diffractive scattering. Although the  dimuon channel is the less copious one, with only about 
840 events produced, almost 41\% of these events are produced by a diffractive process.
This can be understood by recalling our discussions in Sec.~\ref{subsec:cohdiff}.

Finally, we note that a dedicated high-energy run at DUNE has been mooted, to be undertaken after the full period of data collecting for the oscillation analysis. Thanks to the higher energies of the beam, this has the potential to see a significant number of neutrino tridents, provided it can collect enough POTs.  

\subsubsection{$\nu$STORM}
\label{subsubsec:nuSTORM}
In this section we study the trident rates for a possible LAr detector for the proposed 
$\nu$STORM experiment \cite{Soler:2015ada,nuSTORM2017}. The $\nu$STORM facility 
is based on a neutrino factory-like design and has the goal to search for sterile neutrinos and study neutrino nucleus cross sections \cite{Adey:2014rfv}. Although this proposal is in its early days, $\nu$STORM has the potential to make cross section measurements with unprecedented precision. In its current design, $120$-GeV protons are used to produce pions from a fixed target with the pions subsequently decaying into muons and neutrinos. The muons are captured in a storage ring and during repeated passes around the ring they decay to produce neutrinos.
Consequently, the storage ring is an intense source of three types of neutrino
flavours: $\nu_\mu$ from $\pi^+$ and $K^+$ decays, which will be more than $99\%$ of the total flux, $\nu_e$ and $\overline\nu_\mu$ from recirculated muon decays which will comprise less than $1\%$ of the total flux. An important point, however, is that the neutrinos coming from the pion and kaon decays can be separated by event timing from the ones produced by the stored muons. This distinction allows the $\nu_\mu$ flux to be studied almost independently from the $\overline{\nu}_\mu$ and $\nu_e$ flux. In addition, it implies after the initial flash of meson-derived events, that the flux consists of as many electron neutrinos as muon antineutrinos. We will assume a LAr detector for $\nu$STORM at a baseline of 50\,m with 100\,t of fiducial mass with an exposure of $10^{21}$ POT. The neutrino fluxes, assuming 
a central $\mu^+$ momentum of $3.8$~GeV/c in the storage ring, are taken from Ref.~\cite{nuSTORM2017} and are 
shown in Fig.~\ref{fig:LAr}.

In Tab.~\ref{tab:LArrates}, we show the results of our calculations for $\nu$STORM. 
More than $97\%$ of the events from the incoming $\nu_\mu$ are from pion decays and only less 
than $3\%$ from kaon decays. Since we only consider the decay of mesons with positive charges and we expect neutral and wrong charge contamination to be small, we do not have trident events from incoming $\overline\nu_e$.
The total number of mixed flavour, dielectron and dimuon channel events is, respectively,
230, 125 and 20, much less than what can be achieved at the larger neutrino energies available at the DUNE ND. The novel flavour structure of the beam does enhance the contribution of $\nu_e$ induced tridents with respect to the $\pbar{\nu}_\mu$ ones, but this contribution only becomes dominant for the $e^+e^-$ tridents in the muon decay events. Finally, we emphasize that the experimental design parameters for $\nu$STORM are far from definite. Increasing the energy of stored muons and the size of the detector are both viable options which could significantly enhance the rates we present.

%
\subsection{Kinematical Distributions at DUNE ND}
\label{subsec:kine}

\begin{figure}[H]
\centering
\includegraphics[width=\textwidth]{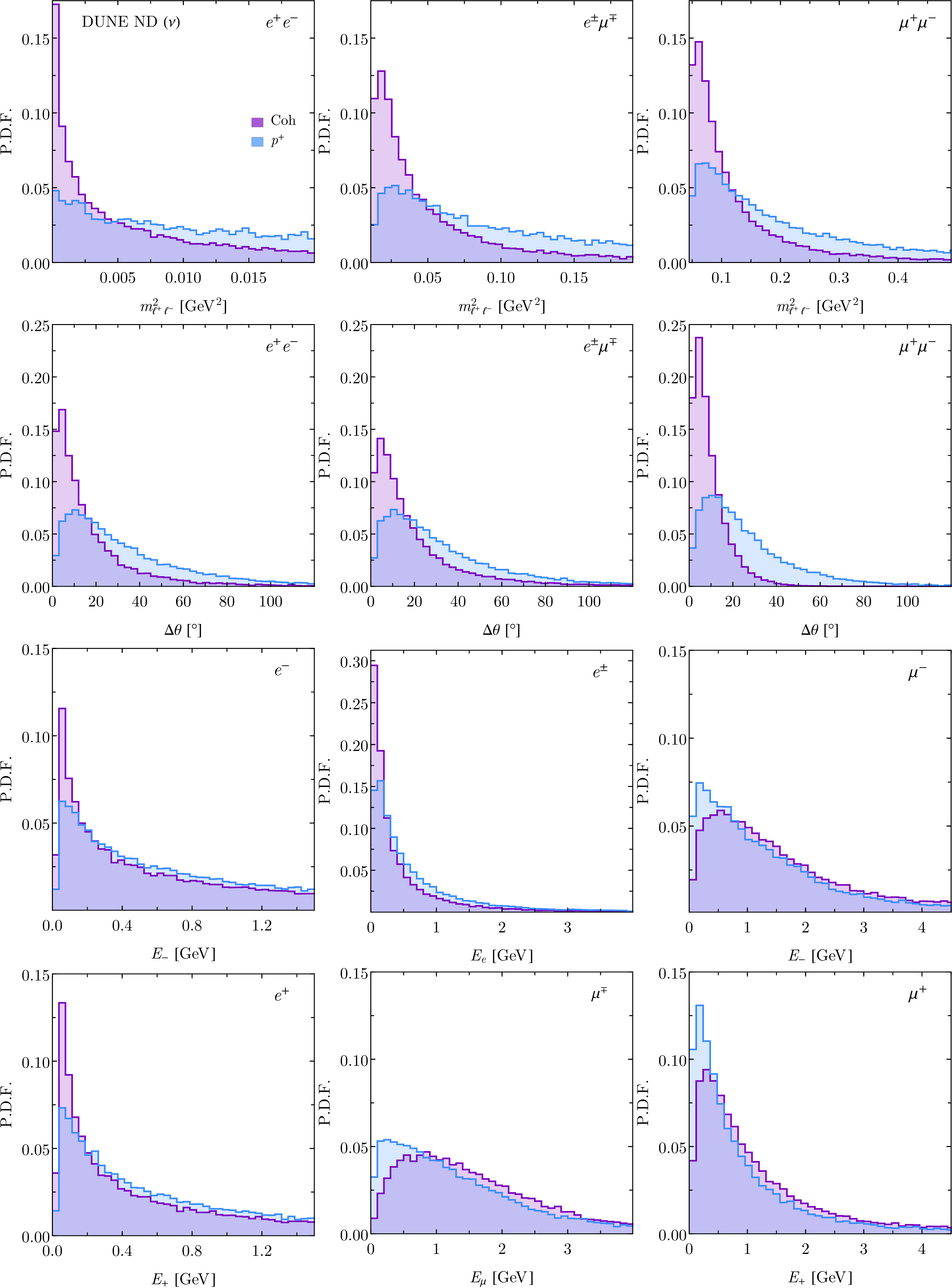}
\caption{Flux convolved neutrino trident production distributions for DUNE ND in neutrino mode. In purple we show the coherent contribution in $^{40}$Ar and in blue the diffractive contribution from protons as targets only (including Pauli blocking). The coherent and diffractive distributions are normalized independently. The relative importance of each contribution as a function of $E_\nu$
can be seen in Fig.~\ref{fig:RatioCDvsT}.
\label{fig:DUNE_ND_dist}}
\end{figure}

In this section we explore the trident signal in more detail, showing some relevant kinematical distributions for coherent and diffractive events. For concreteness, and due to its large number of events, we choose to focus on the DUNE ND, only commenting slightly on the signal at the lower energies of SBN and $\nu$STORM. The observables we calculate are the invariant mass of the charged leptons $m^2_{\ell^+ \ell^-}$, their separation angle $\Delta \theta$ and their individual energies $E_\pm$. The flux convolved distributions of these observables are shown for the DUNE ND in neutrino mode in \reffig{fig:DUNE_ND_dist}. In these plots, we sum all trident channels with a given undistinguishable final-state proportionally to their rates, although $\nu_\mu$ initiated processes always dominate. The coherent and diffractive contributions are shown separately and on the same axes, but we do not worry about their relative normalization. Other potentially interesting quantities are the angle between the cone formed by the two charged leptons and the beam, $\alpha_C$, and the angle of each charged lepton with respect to the beam direction, $\theta_\pm$.  These additional observables are explored in \refapp{app:distributions}. We also report the distributions of the momentum transfer to the hadronic system, $Q^2$. Although this is not a directly measurable quantity, it is a strong discriminant between the coherent and diffractive processes. We do not present the antineutrino distributions here, but they are qualitatively similar.

Perhaps one of the most valuable tools for background suppression in the measurement of the $\mu^+\mu^-$ trident signal at CHARM~II, CCFR and NuTeV \cite{Geiregat:1990gz,Mishra:1991bv,Adams:1998yf} was the smallness of the invariant mass $m^2_{\ell^+ \ell^-}$. This feature, shown here on the top row of \reffig{fig:DUNE_ND_dist}, is also present at lower energies, where the distributions become even more peaked at lower values; although, the diffractive events tend to be have a more uniform distribution in this variable. This is also true for the angular separation $\Delta \theta$, where coherent dimuon tridents tends to be quite collimated, with $90\%$ of events having $\Delta \theta < 20^\circ$, whilst diffractive ones are less so, with only $47\%$ of events surviving the cut. This difference is much less pronounced for mixed and dielectron channels, where only half of our coherent events obey $\Delta \theta < 20^\circ$, when $37\%$ of diffractive events do so.

An interesting feature of same flavour tridents induced by a neutrino (antineutrino) is that the negative (positive) charged lepton tends to be slightly more energetic than its counterpart, whilst for mixed tridents muons tend to carry away most of the energy. These considerations are also reflected in the angular distributions. The most energetic particle is also the more forward one. For instance, in mixed neutrino induced tridents, $\sim 80 \%$ of the $\mu^-$ are expected to be within $10^\circ$ of the beam direction, whilst only $\sim 35 \%$ of their $e^+$ counterparts do so (see \refapp{app:distributions} for additional distributions).

Finally, we mention that detection thresholds can also be important for trident channels with electrons in the final-state. Assuming, for example, a detection threshold for muons and electromagnetic (EM) showers of 30 MeV in LAr, we end up with efficiencies of (99\%, 71\%, 77\%, 86\%) for ($\mu^+ \mu^-$, $e^+ e^-$, $e^+ \mu^-$, $e^- \mu^+$) coherent tridents. These efficiencies become (96\%, 91\%, 93\%, 96\%) for diffractive tridents, dropping for $\mu^+\mu^-$ and increasing for all others. For comparison, at the lower neutrino energies of SBND and assuming the same detection thresholds, the efficiencies for coherent and diffractive tridents are slightly lower, (97\%, 57\%, 67\%, 77\%) and (90\%, 81\%, 85\%, 90\%) respectively.

\subsection{Background Estimates for Neutrino Trident in LAr}
\label{subsec:bck}

The study of any rare process is a struggle against both systematic uncertainties in the event rates and unavoidable background processes. True dilepton signatures are naturally rare in neutrino scattering experiments, but with modest rates of particle misidentification a non-trivial background arises. In this section we estimate the background to trident processes in LAr and its impact on the trident measurement. We perform our analysis only for DUNE ND, in neutrino and antineutrino mode, but our results are expected to be broadly applicable to other LAr detectors. We have generated a sample of $1.1 \times 10^6$ background events using GENIE \cite{Andreopoulos2009} for incident electron and muon flavour neutrinos and antineutrinos. It is worth noting, however, that this event sample will in fact be smaller than the total number of neutrino interactions expected in the DUNE ND. 
Our goal, therefore, will be to demonstrate that with modest analysis cuts background levels can be suppressed significantly such that they become comparable to or smaller than the signals we are looking for. In the absence of events that satisfy our background definition, we argue that the frequency of that type of event is less than one in $1.1\times 10^6$ interactions of the corresponding initial neutrino.  

To account for misreconstruction in the detector, we implement resolutions as a gaussian smear around the true MC energies and angles. We assume relative energy resolutions as $\sigma/E = 15\%/\sqrt{E}$ for $e/\gamma$ showers and protons, and $6\%/\sqrt{E}$ for charged pions and muons. Angular resolutions are assumed to be $1^\circ$ for all particles (proton angles are never smeared in our analysis). The detection thresholds are a crucial part of the analysis, since for many channels one ends up with very soft electrons. We take thresholds to be $30$ MeV for muons and $e/\gamma$ showers kinetic energy, $21$ MeV for protons and $100$ MeV for $\pi^{\pm}$ \cite{DUNECDRvolII}.

\subsubsection{Background Candidates}
\label{subsubsec:misID}
We focus on three final-state charged lepton combinations: $\mu^+\mu^-$, $\mu^\pm e^\mp$ and $e^+e^-$. Genuine production of these states is possible in background processes, but usually rare, deriving from meson resonances or other prompt decays. The majority of the background is expected to be from particle misidentification (misID). We assume that protons can always be identified above threshold and that neutrons leave no detectable signature in the detector. In addition, we require no charge ID capabilities from the detector and assume that the interaction vertex can always be reconstructed. Under these assumptions, we have incorporated three misidentifications which will affect our analysis, and give our naive estimates for their rates in Tab.~\ref{tab:misIDlist}. Any other particle pairs are assumed to be distinguishable from each other when needed.
\renewcommand{\arraystretch}{1.2}
\begin{table}[t]
\centering 
\begin{tabular}{|c c|}
\hline\hline
\bf misID & \bf Rate \\
\hline\hline
$\gamma$ as $e^\pm$ & 0.05 \\
\hline
\multirow{2}{*}{$\gamma$ as $e^+e^-$} & 0.1 (w/ vertex)  \\
 & 1 (no vertex + overlapping)  \\
\hline
$\pi^\pm$ as $\mu^\pm$ & 0.1 \\
\hline\hline
\end{tabular}
\caption{\label{tab:misIDlist} Assumed misID rates for various particles in a LAr detector. We take these values to be constant in energy.}
\end{table}

The requirement of no hadronic activity helps constrain the possible background processes, but one is still left with significant events with invisible hadronic activity and other coherent neutrino-nucleus scatterings. These are then reduced by choosing appropriate cuts on physical observables, exploring the discrepancies between our signal and the background. In our GENIE analysis, we include all events that have final-states identical to trident, or that could be interpreted as a trident final-state considering our proposed misID scenarios. Our dominant sources of background for $\mu^+ \mu^-$ tridents are $\nu_\mu$-initiated charged-current events with an additional charged pion in the final-state ($\nu_\mu$CC$1\pi^\pm$). For $e^+e^-$ tridents, the most important processes are neutral current scattering with a $\pi^0$ (NC$\pi^0$), while for mixed $e^\pm \mu^\mp$ tridents, the $\nu_\mu$-initiated charged-current events with a final-state $\pi^0$ (CC$\pi^0$) dominate the backgrounds. In each case, the pion is misidentified to mimic the true trident final-state. Other relevant topologies include charm production, CC$\gamma$ and $\nu_e$CC$\pi^\pm$. For a detailed discussion of these backgrounds processes we refer the reader to \refapp{app:backgrounds}.

\subsubsection{\label{sec:DUNE_bg_rates}Estimates for the DUNE ND}

In this section we provide estimates for the total background for each trident final-state for the DUNE ND. The number of total inclusive CC interactions in the 50 t detector due to neutrinos of all flavours is calculated to be $5.18 \times 10^8$. We scale our background event numbers to match this, and argue that one has to reach suppressions of order $10^{-6} - 10^{-5}$ to have a chance to observe trident events. Whenever our cuts remove all background events from our sample, we assume the true background rate is one event per $1.1\times10^6$ $\nu$ interactions and scale it to the appropriate number of events in the ND, applying the misID rate whenever relevant. Within our framework, this provides a conservative estimate as the true background is expected to be smaller.

\begin{figure}[t]
\centering
\includegraphics[width = 0.75\textwidth]{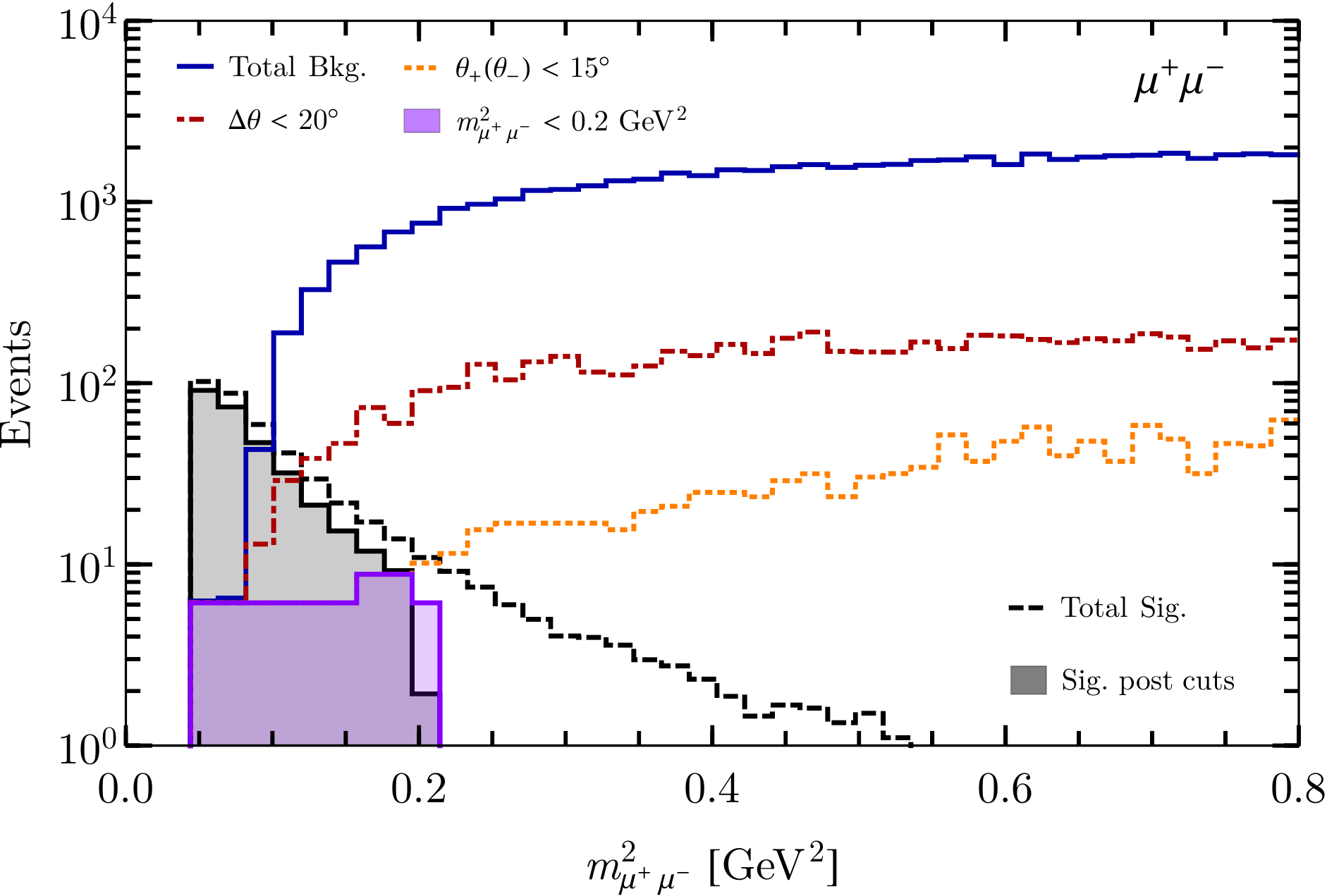}
\caption{Signal and background distributions in invariant mass. The total background events (blue) include the misID rates in table \reftab{tab:misIDlist}. We apply consecutive cuts on the background, starting with cuts on the separation angle $\Delta \theta$ (red), both charged lepton angles to the beamline ($\theta_+$ and $\theta_-$) (orange) and the invariant mass $m^2_{\mu^+ \mu^-}$ . We show the signal samples before and after all the cuts in dashed black and filled black, respectively. \label{fig:bkg_flow}}
\end{figure}

Our estimates are shown in \reftab{tab:DUNE_ND_NU_BG}. We start with the total number of background candidates $\rm N_B^{\mathrm{misID}}$, using only the naive misID rates shown in \reftab{tab:misIDlist}. These are much larger than the trident rates we expect, by at least 2 orders of magnitude. Next, we veto any hadronic activity at the interaction vertex, obtaining $\rm N_B^{\mathrm{had}}$. We emphasize that this veto also affects the diffractive tridents in a non-trivial way, and therefore we remain agnostic about the hadronic signature of these. 
Finally, one can look at the kinematical distributions of coherent trident in \refsec{subsec:kine} and try to estimate optimal one dimensional cuts for the DUNE ND based on the kinematics of the final-state charged leptons. This is a simple way to explore the striking differences between the peaked nature of our signal and the smoother background. In a real experimental setting it is desirable to have optimization methods for isolating signal from background, preferably with a multivariate analyses. However, even in our simple analysis, cutting on the small angles to the beamline and the low invariant masses of our trident signal can achieve the desired background suppressions. For the $\mu^+\mu^-$ tridents we show the effect of our cuts in \reffig{fig:bkg_flow}. The cuts are defined to be $m^2_{\mu^+ \mu^-} < 0.2 \ \mathrm{GeV}^2$, $\Delta \theta < 20^\circ$, $\theta_\pm < 15^\circ$. The kinematics is very similar in the other trident channels, with slightly less forward distributions for electrons. For the $e^+ e^-$ channel we take  $m^2_{e^+ e^-} < 0.1 \ \mathrm{GeV}^2$, $\Delta \theta < 40^\circ$ and $\theta_\pm < 20^\circ$. 
The asymmetry between the positive and negative charged leptons is visible in the distributions, where the latter tends to be more energetic. This feature was not explored in our cuts, as it is not significant enough to further improve background discrimination. In the mixed flavour tridents, however, one sees a much more pronounced asymmetry. The muon tends to carry most of the energy and be more forward than the electron, which can make the search for this channel more challenging due to the softness of the electron in the high energy event. Nevertheless, the low invariant masses and forward profiles can still serve as powerful tool for background discrimination, provided the event can be well reconstructed. We assume that is the case here and use the following cuts on the background:  $m^2_{e^\pm \mu^\mp} < 0.1 \  \mathrm{GeV}^2$, $\Delta \theta < 20^\circ$, $\theta_e < 40^\circ$ and $\theta_\mu < 20^\circ$. When performing kinematical cuts, we also include the effects of detection thresholds after smearing. For a discussion on the impact of these thresholds on the trident signal see \refsec{subsec:kine}. 

The resulting signal efficiencies due to our cuts and thresholds are shown in the last two columns of \reftab{tab:DUNE_ND_NU_BG}. One can see that these are all $ \approx 50\%$ or greater for our coherent samples, whilst all background numbers remain much below the trident signal. The diffractive samples are also somewhat more affected by our cuts than the coherent ones. If one is worried about the contamination of coherent events by diffractive ones, then the kinematics of the charged leptons alone can help reduce this, independently of the hadronic energy deposition of the events. For instance, in the case where all $\mu^+\mu^-$ diffractive events appear with no hadronic signature, then after our cuts the diffractive contribution is reduced from $41\%$ to $15\%$ of the total trident signal. This reduction is, however, also subject to large uncertainties coming from nuclear effects. In summary, the set of results above are encouraging, suggesting that the signal of coherent-like trident production is sufficiently unique to allow for its search at near detectors despite naively large backgrounds. 

\begin{table}[t]	
	\begin{center}
    \resizebox{\textwidth}{!}{
		\begin{tabular}{|clllll|}
		\hline \hline
		\bf Channel& $\bf N^{\mathrm{misID}}_{\mathrm{B}} / N_{\mathrm{CC}}$        & $ \bf N^{\mathrm{had}}_{\mathrm{B}} / N_{\mathrm{CC}}$ & $\bf N^{\mathrm{kin}}_{\mathrm{B}} / N_{\mathrm{CC}}$ & ${\epsilon_{\mathrm{sig}}^{\mathrm{coh}}}$ &
${\epsilon_{\mathrm{sig}}^{\mathrm{dif}}}$ \footnotemark\\	\hline \hline
		$e^{\pm}\mu^{\mp}$ & $1.67\  (1.62) \times 10^{-4}$ & $2.68\  (4.31) \times 10^{-5}$ & $4.40 \ (3.17) \times 10^{-7}$ & $ 0.61 \ (0.61)$ & $ 0.39 \ (0.39)$\\
		$e^+e^-$ & $2.83 \ (4.19)\times 10^{-4}$ & $1.30 \ (2.41) \times 10^{-4}$ &  $6.54 \ (14.1) \times 10^{-6}$ & $ 0.48 \ (0.47)$ & $ 0.21 \ (0.21)$\\
		$\mu^+\mu^-$ & $2.66 \ (2.73)\times 10^{-3}$ & $10.4 \ (9.75)\times 10^{-4}$ & $3.36 \ (3.10)\times10^{-8}$ & $0.66 \ (0.67)$ & $0.17 \ (0.16)$\\\hline\hline
		\end{tabular}
    }
	\end{center}
	\caption{\label{tab:DUNE_ND_NU_BG} Reduction of backgrounds at the DUNE ND in neutrino (antineutrino) mode and its impact on the signal for each distinguishable trident final-state. $\mathbf{N^{\mathrm{misID}}_{\mathrm{B}}}$ stands for total backgrounds to trident after only applying misID rates, $\mathbf{N^{\mathrm{had}}_{\mathrm{B}}}$ are the backgrounds after the hadronic veto, and $\mathbf{N^{\mathrm{kin}}_{\mathrm{B}}}$ reduce the latter with detection thresholds and kinematical cuts (see text for the cuts chosen). These quantities are normalized to the total number of CC interactions in the ND $\mathbf{N_{\mathrm{CC}}}$ (flavour inclusive). We also show the impact of our detection thresholds and kinematical cuts on the trident signal via efficiencies for coherent only ($\epsilon_{\mathrm{sig}}^{\mathrm{coh}}$) and diffractive only samples ($\epsilon_{\mathrm{sig}}^{\mathrm{dif}}$). We do not cut on the hadronic activity of diffractive events.}
\end{table}

\footnotetext{Despite the fact that many diffractive events will likely deposit hadronic energy in the detector, we quote the efficiency of our cuts on diffractive events with no assumptions on their hadronic signature.}

Finally, we comment on some of the limitations of our analysis. The low rate of trident events calls for a more careful evaluation of other subdominant processes that could be easily be overlooked. For channels involving electrons, it is possible that de-excitation photons and internal bremsstrahlung become a source of background, as these also produce very soft EM showers, none of which are implemented in GENIE. The question of reconstruction of these soft EM showers, accompanied either by a high energy muon or by another soft EM shower also would have to be addressed, especially in the latter case where a trigger for these soft events would have to be in place. A more complete analysis is also needed for treating the decay products of charged pions and muons produced in neutrino interactions, as well as rare meson decay channels (like the Dalitz decay of neutral pions $\pi^0 \to \gamma e^+ e^-$). Cosmic ray events are not expected to be a problem due to the requirement of a vertex and a correlation with the beam for trident events. Perhaps even more exotic processes with three final-state charged leptons, like the radiative trimuon production \cite{Smith:1977nx}, could also behave as a background when a single particle is undetected. We are not aware of any estimates for the rate of processes of the type $\nu_{\alpha} (\overline{\nu}_{\alpha}) + \mathcal{H} \to  \ell_\alpha^- (\ell_\alpha^+) + \ell_\beta^+ + \ell_{\beta}^- + \mathcal{H^\prime}$ at the DUNE ND, but note that its rate is comparable to neutrino trident production at energies above $30$ GeV \cite{Albright:1978mg}. Improvements on our analysis should come from the collaboration's sophisticated simulations, allowing for a better quantification of hadronic activity, more realistic misID rates and more accurate detector responses.

\section{Trident Events in Other Near Detector Facilities}

\label{sec:others}
The search for neutrino trident production events certainly benefits from the capabilities of LAr technologies but need not be limited to it. In this section we study neutrino trident production rates at non-LAr experiments which have finished data taking or are still running: the on-axis near detector of T2K (INGRID), the near detectors of MINOS and NO$\nu$A and the MINER$\nu$A experiment. We calculate the total number of trident events as in Eq.~(\ref{eq:nevents}), taking into account the fact that some detectors are made of composite material. We summarize in Tab.~\ref{tab:others} the details of all non-LAr detectors considered in this section. We limit ourselves to a discussion of the total rates in the fiducial volume, but remark that a careful consideration of each detector is needed in order to assess their true potential to detect a trident signal. For instance, requirements about low energy EM shower reconstruction, hadronic activity measurements and event containment would have to be met to a good degree in order for the detector to be competitive. 

\begin{table}[t]
\begin{center}
\scalebox{0.72}{
\begin{tabular}{|c|c|c|c|c|c|}
\hline\hline
\bf Experiment& \bf Material & \bf Baseline (m) & \bf Exposure (POT) & \bf Fiducial Mass (t) & \bf $\mathbf{E_\nu}$ (GeV)\\\hline
INGRID  & Fe & 280  & $3.9\times10^{21}$ [$10^{22}$] T2K-I [T2K-II] & 99.4 & $0-4$ \\\hline
MINOS[+] & Fe and C &1040  & $10.56(3.36)[9.69]\times 10^{20}$  & 28.6& $0-20$ \\\hline
NO$\nu$A & ${\rm{C_2 H_3 Cl}} $ and $ {\rm{C H_2}}$  & 1000 &$8.85(6.9)~[36(36) ]\times10^{20}$ [NO$\nu$A-II] &231 & $0-20$ \\\hline
MINER$\nu$A & ${\rm{CH,H_2O}}, {\rm{Fe,Pb,C}}$ & 1035  & $12(12)\times 10^{20}$ &7.98 & $0-20$ \\\hline\hline
\end{tabular}}
\end{center}
\caption{\label{tab:others} Summary of the non-LAr detector set-up and values used in our calculations. The POT numbers are given for a neutrino (antineutrino) beam. For T2K-I and II 
neutrino and antineutrino beams have the same exposure.}
\end{table}

\subsection{INGRID}
\label{subsec:INGRID}
INGRID, the on-axis near detector of the T2K experiment, is located 280 m from the beam source. It consists of 14 identical iron modules, each with a mass of $7.1$~t, resulting in a total fiducial mass of $99.4$ t~\cite{Abe:2011xv}. The modules are spread over a range of angles between $0^\circ$ and $1.1^\circ$ with respect to the beam axis. The currently approved T2K exposure is $(3.9+3.9)\times 10^{21}$ 
POT in neutrino + antineutrino modes (T2K-I), with the goal to increase it to a total exposure of $(1+1)\times 10^{22}$ POT in the second phase of the experiment (T2K-II) \cite{Abe:2016tez}. Hence we expect approximately $2.6$ times more trident events for T2K-II. 

\begin{figure}[t]
\centering
\includegraphics[width=\textwidth]{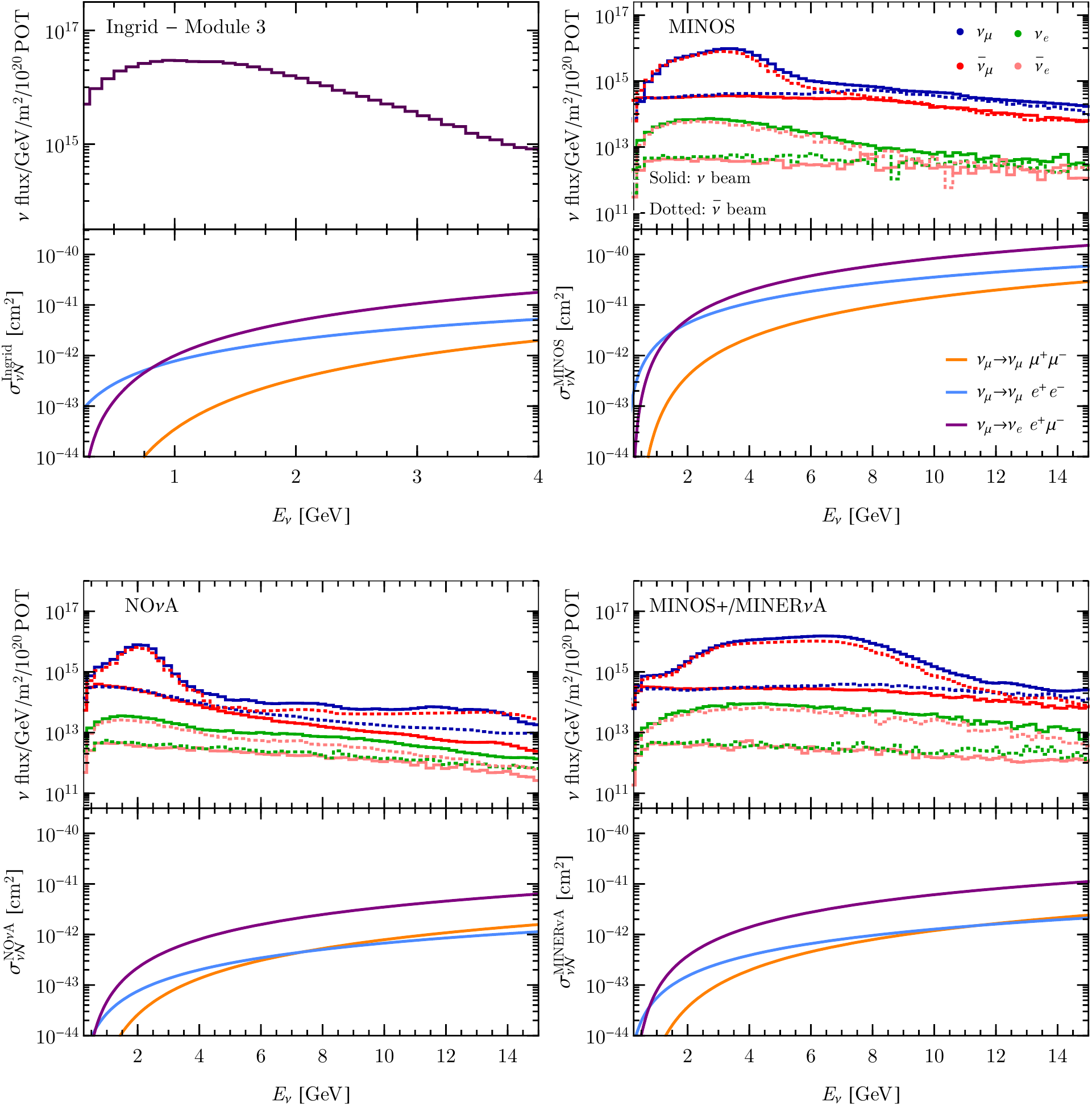}
\caption{\label{fig:others}
Energy distribution of the neutrino fluxes at the position of the detector (top plot) and 
corresponding total trident production cross sections (bottom plot) for:
INGRID~\cite{Abe:2015biq} (first panel), MINOS ND~\cite{fluxes:nonLAr} (second panel), NO$\nu$A ND\cite{fluxes:nonLAr} (third panel)
and MINER$\nu$A\cite{fluxes:nonLAr} (fourth panel).
The cross sections show here for the composite detectors are normalized by the total number of 
atoms.}
\end{figure}

We use the on-axis neutrino mode flux spectra at the INGRID module-3 from Ref.~\cite{Abe:2015biq}, as shown on the top of the first panel of Fig.~\ref{fig:others}. The flux contribution for each neutrino flavour and  energy range is listed in Table 1 of Ref.~\cite{Abe:2015biq}. The total neutrino flux flavour composition at module-3 is 92.5\% $\nu_\mu$, 5.8\% $\overline\nu_\mu$,  1.5\% $\nu_e$ and 0.2\%  $\overline\nu_e$. We assume here that the fluxes at the other 13 modules are the same as at module-3. Although this is not exactly correct it should provide a reasonable estimate of the total rate.

Under these assumptions we show the total number of trident events we calculated for INGRID in 
the first (second) column of Tab.~\ref{tab:otherrates} for T2K-I (T2K-II) exposure.
We predict about 660 (1700) events for the mixed, 300 (770) events for the dielectron and 50 (130) 
events for the dimuon channel for T2K-I (T2K-II). These numbers, although less than 
those expected at the DUNE ND, are already very significant and worth further consideration. We expect, however, that the main challenge will be the reconstruction of final state electrons in these iron detectors.

\begin{table}[t]
\begin{center}
\scalebox{0.8}{
\begin{tabular}{|cccccccc|}
\hline\hline
		\bf Channel & \bf T2K-I& \bf T2K-II & \bf MINOS & \bf MINOS+ & \bf NO$\nu$A-I & \bf NO$\nu$A-II & \bf MINER$\nu$A \\ \hline \hline
		$\nu_\mu\to\nu_e e^+ \mu^-$& $538$ &$1379$ &$179~(25)$&$688$ &$71~(14)$ &$291~(73)$& $140~(13)$ \\
        &$92$ &$235$ &$37~(5)$ &$142$ &$21~(4)$ &$86~(21)$ & $53~(5)$\\\hline
        $\overline\nu_\mu\to\overline\nu_e e^- \mu^+$&$23$ &$58$ &$42~(31)$&$38$ &$10~(57)$ &$41~(296)$ & $8~(89)$\\
        &$4$ &$10$ &$9~(6)$&$8$ &$3~(17)$ &$12~(88)$& $3~(34)$\\\hline
	$\nu_e\to\nu_\mu e^- \mu^+$& $2$ &$6$ &$1~(0.2)$&$4$ &$2~(0.5)$&$8~(3)$ & $1~(0.09)$\\
        &$0.5$ &$1$ &$0.4~(0.05)$ &$1$&$0.9~(0.2)$ &$4~(1)$& $0.4~(0.05)$\\\hline
	$\overline\nu_e\to\overline\nu_\mu e^+ \mu^-$& $0.2$ &$0.6$ &$0.4~(0.3)$&$0.4$ &$0.5~(0.9)$ &$2~(5)$& $0.06~(0.5)$\\
    &$0.06$ &$0.2$ &$0.1~(0.08)$ &$0.1$&$0.2~(0.4)$ &$0.8~(2)$& $0.03~(0.3)$\\\hline
    \hline\hline
    {$\rm{Total} \ e^\pm \mu^\mp$}& $563$ &$1444$ &$222~(56)$ &$730$&$83~(72)$ &$340~(374)$& $149~(102)$ \\
    &$96$ &$246$ &$46~(11)$&$151$ &$25~(22)$ &$102~(114)$& $56~(39)$\\\hline
    \hline
    		$\nu_\mu\to\nu_\mu e^+ e^-$& $257$ &$659$ &$48~(5)$ &$44$&$22~(3)$ &$90~(16)$& $35~(3)$ \\
        &$22$ &$58$ &$6~(0.8)$ &$6$&$3~(0.6)$ &$00$& $9~(0.8)$\\\hline
        $\overline\nu_\mu\to\overline\nu_\mu e^- e^+$& $10$ &$26$ &$9~(8)$&$9$ &$2~(16)$ &$8~(83)$& $2~(23)$\\
        &$0.9$ &$2$ &$1~(1)$ &$1$&$0.4~(3)$ &$2~(15)$& $0.5~(6)$\\\hline
	$\nu_e\to\nu_e e^- e^+$&$9$ &$24$ &$3~(0.3)$ &$8$&$3~(0.9)$ &$12~(5)$& $2~(0.2)$ \\
    &$0.8$ &$2$ &$0.4~(0.05)$ &$1$&$0.7~(0.2)$ &$3~(1)$& $0.4~(0.04)$\\\hline
	$\overline\nu_e\to\overline\nu_e e^+ e^-$&$0.9$ &$2$ &$0.7~(0.6)$&$0.7$ &$0.8~(2)$ &$3~(10)$& $0.1~(0.9)$ \\
    &$0.08$ &$0.2$ &$0.1~(0.08)$ &$0.1$&$0.2~(0.3)$ &$0.8~(1)$& $0.03~(0.2)$\\\hline
    \hline\hline
    ${\rm{Total}}\  e^+ e^-$& $277$ &$711$ &$61~(15)$ &$62$&$29~(22)$ &$119~(114)$& $39~(27)$\\
    &$24$ &$62$ &$9~(2)$ &$8$&$4~(4)$ &$16~(21)$& $10~(7)$\\\hline
    \hline    
    		$\nu_\mu\to\nu_\mu \mu^+ \mu^-$& $29$ &$73$ &$21~(3)$ &$81$&$7~(2)$ &$28~(11)$& $17~(2)$ \\
        &$20$ &$52$ &$12~(2)$ &$46$&$7~(2)$ &$29~(10)$& $17~(2)$\\\hline
        $\overline\nu_\mu\to\overline\nu_\mu \mu^- \mu^+$& $1$ &$3$ &$5~(3)$&$5$ &$1~(7)$ &$4~(35)$& $1~(11)$\\
        &$0.9$ &$2$ &$3~(2)$ &$3$&$1~(6)$ &$4~(30)$& $1~(11)$\\\hline
 $\nu_e\to\nu_e \mu^+ \mu^-$    &$0.09$ &$0.2$ &$0.09~(0.01)$&$0.3$ &$0.1~(0.04)$ &$0.4~(0.2)$& $0.06~(0.007)$\\  
 &$0.05$ &$0.1$ &$0.04~(0.006)$ &$0.1$&$0.1~(0.03)$ &$0.4~(0.1)$& $0.05~(0.005)$\\\hline
        $\overline\nu_e\to\overline\nu_e \mu^+ \mu^-$&$0.01$ &$0.03$ &$0.03~(0.02)$ &$0.03$&$0.04~(0.06)$&$0.2~(0.3)$ & $0.004~(0.03)$\\
        &$0.006$ &$0.01$ &$0.01~(0.009)$ &$0.01$&$~0.03(0.05)$  &$0.1~(0.3)$ & $0.003~(0.03)$\\\hline
        \hline \hline
    ${\rm{Total}} \ \mu^+ \mu^-$ &$30$ &$76$ &$26~(6)$&$86$ &$9~(9)$ &$37~(47)$ & $18~(13)$ \\
    &$21$ &$54$ &$15~(3)$ &$49$&$8~(8)$ &$34~(36)$ & $18~(13)$\\
    \hline\hline
\end{tabular}}
\end{center}
\caption{\label{tab:otherrates}Total number of \textbf{coherent} (top row) and \textbf{diffractive} (bottom row) trident events expected at different non-LAr detectors for each channel. The numbers in parentheses are for the antineutrino running mode, when present. These calculations consider a detection efficiency of 100\%.}
\end{table}

\subsection{MINOS/MINOS+ Near Detector}
\label{subsec:MINOS}
The MINOS near detector is a magnetized, coarse-grained tracking calorimeter, made primarily of steel and plastic scintillator. Placed $1.04$~km away from the NuMI target at Fermilab \cite{Aliaga:2016oaz}, it weighs $980$~t and is similar to the far detector in design. In our analysis, we assume a similar fiducial volume cut to the standard $\nu_\mu$ CC analyses, namely a fiducial mass of $28.6$~t made
of $80\%$ of iron and $20\%$ of carbon \cite{Boehm:2009zz}.

The experiment ran from 2005 till 2012 in the low energy (LE) configuration of the NuMI beam ($E_\nu^{\rm peak} \approx 3$ GeV) and collected $10.56\times 10^{20} ~(3.36\times 10^{20})$ POT in the neutrino (antineutrino) beam \cite{Aurisano}. The successor to MINOS, MINOS+, ran with the same detectors subjected to the medium energy (ME) configuration of the NuMI beam ($E_\nu^{\rm peak} \approx 7$ GeV) from 2013 to 2016, and has collected $9.69\times 10^{20}$ POT in the neutrino mode.
To calculate the trident event rates we use the fluxes taken from Ref.~\cite{fluxes:nonLAr}. 
The flavour composition at MINOS ND is 89\% (18\%) $\nu_\mu$ and 10\% (81\%) $\overline \nu_\mu$ for the neutrino (antineutrino) beam and about 1\%  $\nu_e+\overline\nu_e$ for either beam mode. We assume that the MINOS+ neutrino flux is identical to the one at the MINER$\nu$A experiment (see section \ref{subsec:MINERvA}).
These fluxes and total trident production cross sections are shown on the second  panel of Fig.~\ref{fig:others}.

Due to the multi-component material of the detector, the corresponding cross sections that 
enter in Eq.~(\ref{eq:nevents}) are:
\begin{equation}
\sigma_{\rm \nu X}^{\rm{MINOS}}= \sum_{i=\rm Fe,C} f_i \, \sigma_{\rm \nu X}^{i}\,,
\end{equation}
where $f_i$ is the number of nuclei $i$ over the total number of nuclei in the detector.
As a reference, the weighted cross sections, normalized by the total number of atoms, is 
also shown in Fig.~\ref{fig:others}. 

We report the total number of trident events for MINOS ND in Tab.~\ref{tab:otherrates}. 
Although the cross section for iron is about two times 
larger than for argon and the neutrino fluxes similar, the number of trident events at MINOS ND is much smaller than the expected one at DUNE ND due to a lower exposure and fiducial mass.
We predict that about 270 (68) mixed, 70 (17) dielectron and 40 (9)   dimuon trident events
were produced at this detector with the neutrino (antineutrino) LE NuMI beam. 
The rates are expected to be larger for MINOS+, as it benefits from the larger energies of the ME NuMI beam configuration and has similar number of POT to MINOS in neutrino mode. In total, we predict about 880 mixed, 70 dielectron and 135 dimuon trident events.

The stringent cut on the fiducial volume assumed here implies a reduction from the $980$~t near detector bulk mass to $28.6$~t. This cut can be relaxed, depending on the signature considered, and may significantly enhance the rates we quote. A careful analysis of trident signatures outside the fiducial volume would be necessary, but we point out that our rates can increase by at most a factor of $\approx30$.

\subsection{NO$\nu$A Near Detector}
\label{subsec:NOvA}

The NO$\nu$A near detector is a fine grained low-Z liquid-scintillator detector placed off-axis from the NuMI beam at a distance of $1$~km. Its total mass is $330$~t, with almost $70$\% of it active mass ($231$~t). In this analysis we assume all of this active mass to also be fiducial. The detector is mainly made of 70\% mineral oil (CH$_2$) and 30\% of PVC (${\rm{C_2H_3Cl}}$) \cite{Wang:Biao}. A total exposure of $8.85 \, (6.9)\times10^{20}$ POT has been collected in the neutrino (antineutrino) beam mode prior to 2018~\cite{sanchez_mayly_2018_1286758}. 

The NO$\nu$A ND neutrino fluxes (taken from Ref. \cite{fluxes:nonLAr}) peak at slightly lower energies than the MINOS or MINER$\nu$A ones, $E_\nu^{\rm peak} \approx 2$ GeV, and are shown in the third panel of Fig. \ref{fig:others}. The flavour composition 
is 91\% (11\%) $\nu_\mu$ and 8\% (88\%) $\overline \nu_\mu$ in the neutrino (antineutrino) 
mode and about 1\% $\nu_e+\overline\nu_e$ in each mode.

Here the cross sections entering in Eq.~(\ref{eq:nevents}) are calculated as
\begin{eqnarray}
\sigma^{\rm{NO\nu A}}_{\rm \nu X}=\sum_{i=\rm C, Cl, H} f_i \sigma_{\rm \nu X}^{i} \,,
\end{eqnarray}
where $f_i$ is the number of nuclei $i$ over the total number of nuclei in the detector.
As a reference, the weighted cross sections, normalized by the total number of atoms, is shown 
in Fig.~\ref{fig:others}. 

In Tab.~\ref{tab:otherrates} we show our predictions for the number of trident events at 
NO$\nu$A ND. 
Comparing NO$\nu$A and MINOS, we see that while NO$\nu$A ND has a fiducial mass 
almost 8 times larger, the flux times total cross section at MINOS ND is at least two orders of magnitude larger than at NO$\nu$A ND, especially above 4 GeV (see Fig. \ref{fig:others}), making the rates at MINOS ND larger than the rates at NO$\nu$A ND. 

NO$\nu$A is planning to collect a total exposure of $36\,(36)\times10^{20}$ POT in the neutrino (antineutrino) mode (NO$\nu$A-II) \cite{sanchez_mayly_2018_1286758,NovaII}, making the expected rates almost $4.1 (5.2)$ times larger (shown in Tab.~\ref{tab:otherrates}). In this case the expected dimuons and mixed events at MINOS+ would be at least two times larger than NO$\nu$A-II. On the other hand, for NO$\nu$A-II there will be two times more dielectron events given the much higher exposure. 

\subsection{MINER$\nu$A}
\label{subsec:MINERvA}

The multi-component MINER$\nu$A detector was mainly designed to measure neutrino and antineutrino interaction cross sections with different nuclei in the 1-20 GeV range of energy~\cite{MINERvA:2017}. The detector is located at $1.035$~km from the NuMI target. We assume a fiducial mass of about $8$~t, with a composition of 75\% CH, 9\% Pb, 8\% Fe, 6\% H$_2$O and $2\%$ C. The experiment has collected $12\times10^{20}$ POT in the neutrino mode and is planning to reach the same exposure in the antineutrino mode by 2019, both using the medium energy flux of NuMI beam configuration (shown in fourth panel of Fig.~\ref{fig:others}). We do not include the low energy runs, as these have lower number of POT and lower neutrino energies. The neutrino (antineutrino) beam is composed of 95\% (7\%) $\nu_\mu$ and 
4\% (92\%) $\overline \nu_\mu$, both beams have  about $1\%$ of $\nu_e+\overline\nu_e$.

For MINER$\nu$A the cross sections  in Eq.~(\ref{eq:nevents}) are calculated as
\begin{eqnarray}
\sigma^{\rm MINER \nu A}_{\rm \nu X}=\sum_{i=\rm C, Cl, H, Pb, Fe,O} f_i \, \sigma_{\rm \nu X}^{i} \,,
\end{eqnarray}
where $f_i$ is the number of nuclei $i$ over the total number of nuclei in the detector.
As a reference, the weighted cross sections, normalized by the total number of atoms, is shown in
Fig.~\ref{fig:others}. 

The total number of trident events we estimate for MINER$\nu$A are listed in Tab.~\ref{tab:otherrates}. As expected, these are lower than MINOS+, as the latter has a larger fiducial mass. MINER$\nu$A, however, benefits from its fine grained technology and its dedicated design for cross section measurements.

\section{Conclusions}

\label{sec:conc}
Neutrino trident events are predicted by the SM, however, only $\overline{\nu}_\mu$ initiated dimuon tridents have been observed in small numbers, typically fewer than 100 events. This will change in the near future thanks to the current and future generations of precision neutrino scattering and oscillation experiments, which incorporate state-of-the-art detectors located at short distances from intense neutrino sources. 
In this work we 
discuss the calculation of the neutrino trident cross section for all flavours and hadronic targets, and provide estimates for the number and distributions of events at 9 current or future neutrino detectors: five detectors based on the new LAr technology 
(SBND, $\mu$BooNE, ICARUS, DUNE ND and $\nu$STORM ND) as well as four more conventional detectors (INGRID, MINOS ND, NO$\nu$A ND and MINER$\nu$A).

We have stressed the need for a full four-body phase space calculation of the trident cross sections without using the EPA. This approximation has been employed in recent calculations and can lead to overestimations of the cross section by 200\% or more at the peak neutrino energies relevant for many accelerator neutrino experiments.
Moreover, we show why the EPA is not applicable for computing trident cross sections, and provide the first quantitative assessment of this breakdown for coherent and diffractive hadronic regimes. 
We find that the breakdown of the approximation is most severe for processes with electrons in the final-state and for diffractive scattering of all final state flavours. 
For coherent dimuon production, the approximation can give a reasonable result at large neutrino energies. This is due to the nuclear form factors that serendipitously suppress those regions of phase space where the EPA is least applicable. We also demonstrated that the best results in this channel are achieved when applying artificial cuts to the phase space.
However, even in this case, at energies relevant for the above experiments, the EPA can artificially suppress the coherent scattering contribution and increase the diffractive one giving rise to an incorrect rate and distributions of observable quantities. 
For instance, the invariant mass of the charged lepton pair $m^2_{\ell \ell}$ and their angular separation $\Delta \theta$ are more uniformly distributed for diffractive than for coherent trident scattering. Using the correct distributions is crucial to correctly disentangle the signal from the background by cutting on these powerful discriminators.

Our calculations show that DUNE ND is the future detector with the highest neutrino trident statistics, more than 6500 mixed events, 18\% produced by diffractive scattering, more than 2000 dielectron events, 12\%  produced by diffractive scattering and about 840 dimuon events,
almost 41\% of those produced by a diffractive process. Making use of our efficiencies (see \reftab{tab:DUNE_ND_NU_BG}), assuming an ideal background suppression and neglecting systematic uncertainties, we quote the statistical uncertainty on the coherent-like flux averaged cross section for the DUNE ND. We do this for coherent only events and, in brackets, for coherent plus diffractive events, yielding
\[\frac{\delta \langle \sigma^{e^\pm\mu^\mp} \rangle}{\langle \sigma^{e^\pm\mu^\mp} \rangle} =  1.8\% \, (1.6\%), \quad \frac{\delta \langle \sigma^{e^+e^-} \rangle}{\langle \sigma^{e^+e^-} \rangle} =  3.4\% \,(3.3\%) \quad \mathrm{and} \quad \frac{\delta \langle\sigma^{\mu^+\mu^-} \rangle}{\langle \sigma^{\mu^+\mu^-} \rangle} =  5.5\% \,(5.1\%).\]
In this optimistic framework we expect the true statistical uncertainty on coherent-like tridents to lie between the two numbers quoted, depending on how many diffractive events contribute to the coherent-like event sample. This impressive precision would provide unprecedented knowledge of the trident process and the nuclear effects governing the interplay between coherent and diffractive regimes. We emphasize, however, that given these small values for the relative uncertainties, the trident cross section will likely be dominated by systematic uncertainties from detector response and backgrounds which are not modelled here. 

For DUNE ND, we have studied the distribution of observables which could help distinguish trident events from the background. We have estimated the background for each trident channel via a Monte Carlo simulation using GENIE, and identified the dominant contributions arising primarily from particle misidentification.  
We conclude that reaching background rates of the order 
${\cal O}(10^{-6}-10^{-5})$ times the CC rate is necessary to observe trident events at DUNE ND, and given the distinctive kinematic behaviour of the trident signal a simple cut-based GENIE-level analysis suggests that this is an attainable goal in a LAr TPC. 

Existing facilities may also be able to make a neutrino trident measurement at their near detectors. Despite not including reconstruction efficiencies nor an indication of the impact of backgrounds, we find that the largest trident statistics is available at INGRID, the T2K on-axis near detector. We predict about 660 (1700) events for the mixed flavour, 300 (770) events for the dielectron and 50 (130) events for the dimuon channel for T2K-I (T2K-II). The more fine-grained near detector of MINOS and MINOS+ is also expected to have collected a significant numbers of events during its run. As such, the very first measurement of neutrino trident production of mixed and dielectron channels may be at hand.

\acknowledgments
The authors would like to thank TseChun Wang for his involvement during the initial stages of this project.

MH would like to thank Alberto Gago, Jos\'e Antonio Becerra Aguilar and Kate Scholberg for useful discussions regarding the detection of trident events. YFPG and MH would like to thank Gabriel Magill and Ryan Plestid for helpful discussions on the cross section computation considering the EPA. ZT appreciates the useful discussions with Maxim Pospelov and Joachim Kopp. RZF would like to thank Thomas J Carroll for discussions about the MINOS near detector

This work was partially supported by Funda\c{c}\~ao de Amparo \`a
Pesquisa do Estado de S\~ao Paulo (FAPESP) and Conselho Nacional de
Ci\^encia e Tecnologia (CNPq). This project has also received support
from the European Union's Horizon 2020 research and innovation programme under the Marie Sklodowska-Curie grant agreement
No. 690575 (RISE InvisiblesPlus) and No. 674896 (ITN Elusives)
SP and PB are supported by the European Research Council under ERC Grant “NuMass”
(FP7-IDEAS-ERC ERC-CG 617143). SP acknowledges partial support from the Wolfson Foundation and the Royal Society. 

\appendix

\section{\label{app:formfactors}Form Factors}

In the coherent regime, we use a Woods-Saxon (WS) form factor due to its success in reproducing the experimental data \cite{Fricke:1995zz,Jentschura2009}. The WS form factor is the Fourier transform of the nuclear charge distribution, defined as 
\begin{equation}
 \rho(r) = \frac{\rho_0}{1+\exp\left(\dfrac{r - r_0}{a}\right)} \, ,
\end{equation}
where we take $r_0 = 1.126 \, A^{1/3}$ fm and $a = 0.523$ fm. One can then calculate the WS form factor as
\begin{equation}
 F(Q^2) = \frac{1}{\int \rho(r) \, \dd^3r}  \int \rho(r) \, \exp\left( -i \vec{q} \vdot \vec{r}\right) d^3r \, .
\end{equation}
Here we use an analytic expression for the symmetrized Fermi function \cite{Anni1994,Sprung1997} instead of calculating the WS form factor numerically. This symmetrized form is found to agree very well with the full calculation and reads
\begin{equation}
  F(Q^2) =  \frac{3 \pi  a}{r_0^2 + \pi^2 a^2} \frac{\pi a \coth{(\pi Q a)} \sin{(Q r_0)} - r_0 \cos{(Q r_0)} }{Q r_0 \sinh{(\pi Q a)}}\, .
\end{equation}

In the diffractive regime, we work with the functions $H_1^{\rm N}(Q^2)$ and $H_2^{\rm N}(Q^2)$, which depend on the Dirac and Pauli form factors of the nucleon ${\rm N}$ as follows
\begin{equation}
H_1^{\rm N}(Q^2)= |F_1^{\rm N}(Q^2)|^2 - \tau |F_2^{\rm N}(Q^2)|^2\, , \quad \mathrm{and} \quad H_2^{\rm N}(Q^2) = \left| F_1^{\rm N}(Q^2) + F_2^{\rm N}(Q^2)\right|^2 \, ,
\end{equation}
where $\tau = -Q^2/4M^2$. The form factors $F_1^{\rm N}(Q^2)$ and $F_2^{\rm N}(Q^2)$ can be related to the usual Sachs electric $G_\mathrm{E}$ and magnetic $G_{\mathrm{M}}$ form factors. These have a simple dipole parametrization
\begin{align}
G^{\rm N}_E(Q^2) =& F^{\rm N}_1 (Q^2) + \tau F^{\rm N}_2 (Q^2) = \begin{cases}
                                              0,\, &\mathrm{if }\, {\rm N} = n,\\
                                              G_D(Q^2),\, &\mathrm{if }\, {\rm N} = p,
                                              \end{cases} \\
G^{\rm N}_M(Q^2) =& F^{\rm N}_1 (q^2) + F^{\rm N}_2 (Q^2) = \begin{cases}
                                              \mu_n  \, G_D(Q^2),\, &\mathrm{if }\, {\rm N} = n,\\
                                              \mu_p  \, G_D(Q^2),\, &\mathrm{if }\, {\rm N} = p,
                                              \end{cases}
\end{align}
where $\mu_{p,n}$ is the nucleon magnetic moment in units of the nuclear magneton and $G_D(Q^2) = (1 + Q^2/M_V^2)^{-2}$ is a simple dipole form factor with $M_V = 0.84$ GeV.
%

\section{Kinematical Distributions \label{app:distributions}}

\begin{figure}[H]
\centering
\includegraphics[width=0.92\textwidth]{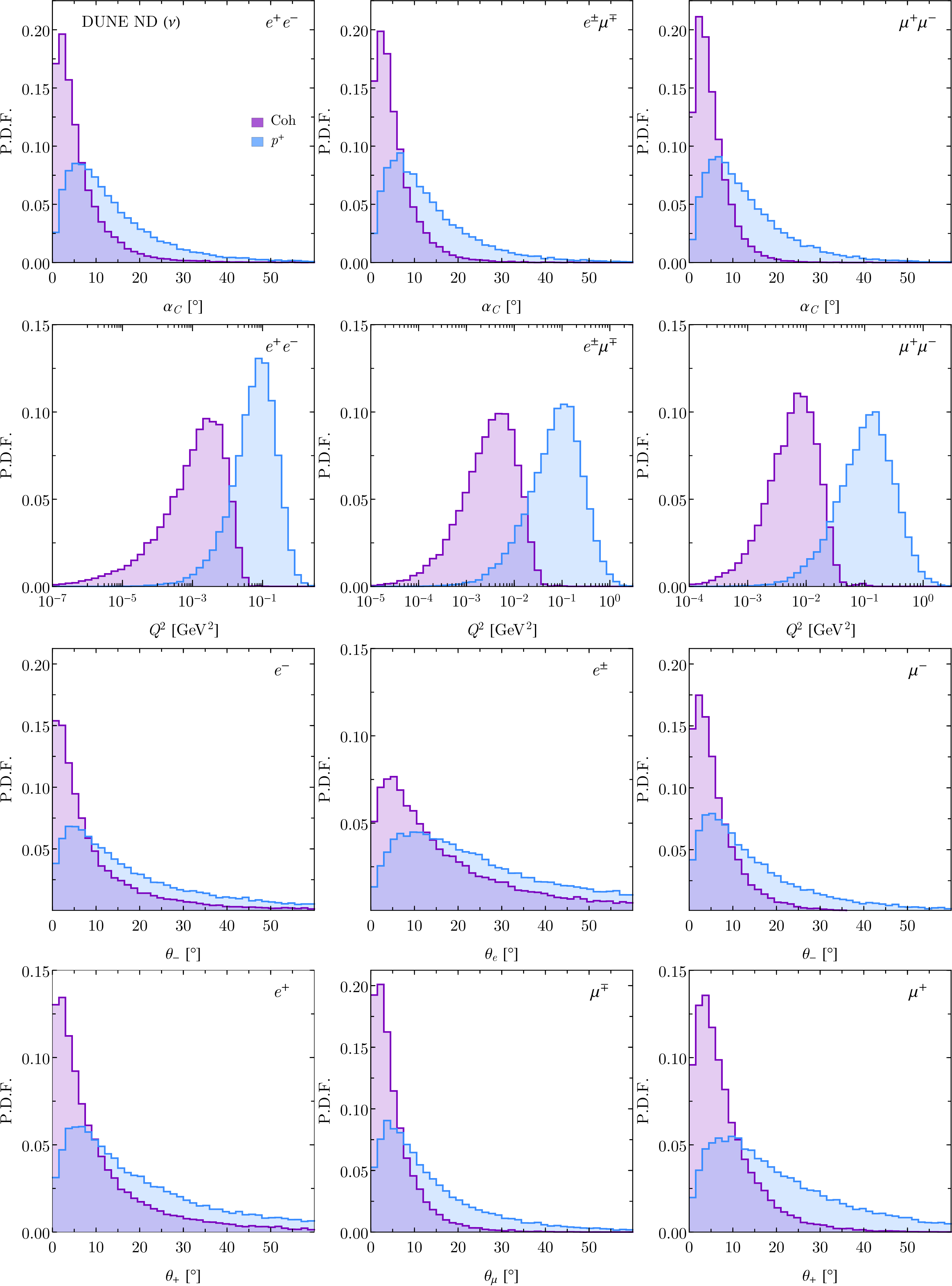}
\caption{Flux convolved neutrino trident production distributions for DUNE ND in neutrino mode in additional variables. In purple we show the coherent contribution in $^{40}$Ar and in blue the diffractive contribution from protons as targets only (including Pauli blocking). The coherent and diffractive distributions are normalized independently. \label{fig:other_dists}}
\end{figure}

In this section, we show additional distributions in different observables for neutrino trident production, also focused on the DUNE ND in neutrino mode. While trident events are generally quite forward going, their angular behaviour is quite interesting. We consider here the angle between the charged lepton cone and the neutrino beam, $\alpha_C$, defined as 
\[
\cos{\alpha_C} = \frac{ (\vec{p}_3 + \vec{p}_4)  \vdot \vec{p}_1}{ |\vec{p}_3 + \vec{p}_4| |\vec{p}_1| }\, ,\]  
and in the individual angle of the charged lepton to the neutrino beam, $\theta$. For same flavour tridents we define $\theta$ for each charge of the visible final-state, whilst for mixed tridents we use their flavour. We also show the distribution in $Q^2 = {-q^2}$, where $q = (P - P^\prime)$, which is of particular interest when considering coherency and the impact of form factors.

\section{Individual Backgrounds}
\label{app:backgrounds}
Here we discuss backgrounds to trident final-states in more detail. We start by motivating our misID rates shown in \reftab{tab:misIDlist}, and then discuss the dominant background processes individually.

In LAr photons can be distinguished from a single electron if their showers start displaced from the vertex (if present). Photons have a conversion length in LAr of around 18 cm, meaning $5$--$10\%$ could be expected to convert quickly enough to hinder electron-photon discrimination by this means if the resolution on the gap is from $1$--$2$ cm \cite{Acciarri:2016sli}. Once pair conversion happens, photons can be distinguished from a single electron purely by $\dd E/\dd x$ measurements in the first 1--2 cm of their showers. Motivated by the success of this method as shown at ArgoNeuT \cite{Acciarri:2016sli} and based on projections for DUNE \cite{Acciarri:2016ooe}, we assume that $5\%$ of photons would be taken as $e^\pm$ with perfect efficiency, without the need for an event vertex. Needless to say that a dedicated study for trident topologies would be necessary for a more complete study. It is worth noting that our remarks concern only the misID of a single photon for a single electron, whilst the distinction between a photon and an overlapping $e^+e^-$ pair without a vertex can be much more challenging. For this reason we take the misID rate between an overlapping $e^+e^-$ pair and a photon to be 1 in the absence of a vertex.  

Charged pions are notorious for faking long muon tracks. We estimate this misID rate as arising from through-going pions, which do not exhibit the decay kink used in their identification. We assume an interaction length of around $1$ m, meaning that about $5\%$ of particles travel $\sim3$ meters and escape the fiducial volume. Assuming that this is the most likely way a pion can spoof a muon, we estimate a naive suppression rate of $10^{-2}$. In a more complete study, it is desirable to explore the length of the muon and pion tracks inside the detector as a function of energy. The length of the contained tracks can also be an important tool for background suppression which we leave to future studies.  

\subsection{Pion Production}

Coherent pion production in its charged ($\nu + A \to \ell^\mp + \pi^\pm + A$) and neutral ($\nu + A \to \nu + \pi^0 + A$) current version is very abundant at GeV energies. The cross section for these processes is modelled in GENIE using a modern version of the Rein-Sehgal model \cite{REIN198329,Rein:2006di}. The charged current version serves mainly as a background to $\mu^+ \mu^-$ tridents, but can also appear as a background for $e^\pm \mu^\mp$ tridents for incoming electron neutrinos or antineutrinos. It has been studied before at MiniBooNE \cite{AguilarArevalo:2010xt}, MINER$\nu$A \cite{Higuera:2014azj,Mislivec:2017qfz}, T2K \cite{Abe:2016fic,Abe:2016aoo}, and for the first time in LAr at ArgoNeuT \cite{Acciarri:2014eit}. This process has a very distict low 4-momentum transfer to the nucleus $|t|$ \cite{Higuera:2014azj}, but a much flatter distribution in invariant mass if compared to trident. The neutral current version of coherent pion production serves as a background to $e^+e^-$ tridents. This process has been studied before by the MiniBooNE \cite{AguilarArevalo:2009ww}, SciBooNE \cite{Kurimoto:2010rc} and in LAr by the ArgoNeuT collaboration \cite{Acciarri:2015ncl}. There are two possibilities for these events to fake an $e^+e^-$ trident: when one of the gammas produced in the $\pi^0$ decay is missed and the other is misIDed for an overlapping $e^+e^-$ pair, and when both photons are each misIDed for a single electron. This signature also comes with low hadronic activity, but for separated visible photons the invariant mass is a natural discriminator, as in the detector $m_{\gamma \gamma} \approx m_{\pi^0}$.

Resonant pion production can also contribute to trident backgrounds in the absence of any reconstructed protons. Resonant pion production can be larger than its coherent counterpart and is modelled in GENIE by the Rein-Sehgal model \cite{Rein:1980wg}. Its CC version was measured by MiniBooNE \cite{AguilarArevalo:2010xt}, K2K \cite{Mariani:2010ez}, MINOS \cite{Adamson:2014pgc}, and MINER$\nu$A \cite{Altinok:2017xua}. In the latter measurement one can clearly see the large number of events with undetected protons. The misIDed photon and the charged lepton invariant mass are once more flatter than the trident ones, allowing for a kinematical discrimination whenever a single photon is undetected. It is worth noting that these are some of the dominant underlying processes for pion production in GENIE, but all events leading to topologies relevant for trident are included in our analysis.
%
\subsection{Charm Production}

Since the first observation of dimuon pairs from charm production in neutrino interaction by the HPWF experiment in 1974 \cite{Benvenuti:1975ru}, a lot has been learned about these processes (see \cite{Lellis:2004yn} for a review) in neutrino experiments. Particularly, the production of charm quarks and their subsequent weak decays into muons or electrons have been identified as a major source of background for early trident searches. At the lower neutrino energies at DUNE, however, this is expected to be a smaller yet non-negligible contribution. From our GENIE samples, we estimate that a charmed state is produced at a rate of around $10^{-4}(N_\text{CC}+N_\text{NC})$. Most of these produce either D mesons, 
$\Lambda_c$ or $\Sigma_c$ baryons. These particles decay in chains, emitting a muon with a branching ratio of around $0.1$, and are always accompanied by pions or other hadronic particles. We therefore expect these rates to be negligible with a hadronic veto, and do not consider them further. We hope, however, that future studies will address these channels in more detail.

\subsection{CC$\gamma$ and NC$\gamma$}

The emission of a single photon alongside a CC process could be a background for $\mu e$ tridents if the photon is misIDed as a single electron. When the photon is produced in a NC event, it can be a background to overlapping $e^+e^-$ tridents. In GENIE, these topologies arise mainly due to resonance radiative decays and from the intra-nuclear processes. For this reason, it usually comes accompanied with extra hadronic activity. For hadronic resonances, we have simulated CC processes in GENIE and estimated
the multiplicities: $0.5\%$ single
photon and $1\%$ double photon emission from CC rates. Radiative photon production from the charged lepton, on the other hand, does not need to come accompanied by hadrons. It is phase space and $\alpha\approx1/137$ suppressed with respect to CCQE rates, and therefore could occur at appreciable rates compared to our signal. This contribution, however, is not included in GENIE and is absent from our samples. The rates of internal photon bremsstrahlung have been estimated before, particularly for T2K where a low-energy photon is an important background for electron
appearance searches \cite{Efrosinin:2009zz}, and as a background to the low energy events at MiniBooNE \cite{Bodek:2007wb}. De-excitation gammas from the struck nuclei can also generate CC$\gamma$ or NC$\gamma$ topologies \cite{PhysRevLett.108.052505}. These contributions for Ar are not included in GENIE, but are expected to come with a distinct energy profile, which can be tagged on.

\bibliographystyle{JHEP}
\bibliography{trident}{}

\end{document}